\documentclass[10pt,journal]{IEEEtran}
\usepackage{cite}
\usepackage{graphicx}
\usepackage{amsmath}
\usepackage{array}
\usepackage{mdwmath}
\usepackage{amssymb}
\usepackage{mdwtab}
\usepackage{url}
\usepackage{stfloats}
\usepackage{amsmath,amsthm}
\usepackage{verbatim}
\usepackage{threeparttable}
\usepackage{color}
\usepackage{ulem}
\usepackage{siunitx}

\usepackage{textcomp}
\ifCLASSOPTIONcompsoc
\usepackage[caption=false,font=normalsize,labelfont=sf,textfont=sf]{subfig}
\else
\usepackage[caption=false,font=footnotesize]{subfig}
\fi

\allowdisplaybreaks[4]

\newtheorem{theorem}{Theorem}

\newtheorem{lemma}{Lemma}

\newcommand{\tabincell}[2]{\begin{tabular}{@{}#1@{}}#2\end{tabular}} 
\graphicspath{{./Figs/}}
\everymath{\displaystyle}

\begin{document}

\title{Performance Analysis for Hybrid {Sub-6GHz-mmWave-THz} Networks with Downlink and Uplink Decoupled Cell Association}

\author{Yunbai Wang, Chen Chen,~\IEEEmembership{Member,~IEEE,} and Xiaoli Chu,~\IEEEmembership{Senior Member,~IEEE}
	
	\thanks{This work was supported in part by the European Unions Horizon 2020
Research and Innovation Program under the Marie Skodowska-Curie Actions
with Grant Agreement No. 778305. \textit{(Corresponding author: Chen Chen.)}
}
 
	\thanks{Yunbai Wang and Xiaoli Chu are with Department of Electronic and Electrical Engineering, the University of Sheffield, Sheffield, S1 4ET, UK. (e-mail: \{ywang286, x.chu\}@sheffield.ac.uk).
		
	Chen Chen is with the School of Electrical Engineering and Computer Science, KTH Royal Institute of Technology, Stockholm, Sweden. (e-mail: chch2@kth.se). }
 }

%\markboth{IEEE TWC}%
%{Shell \MakeLowercase{\textit{et al.}}: Bare Demo of IEEEtran.cls for Journals}

\maketitle
\begin{abstract}
It is expected that 5G/6G networks will exploit sub-6 GHz, millimetre wave (mmWave) and terahertz (THz) frequency bands simultaneously and will increase flexibility in user equipment (UE)-cell association.
In this paper, we introduce a novel stochastic geometry-based framework for the analysis of the signal-to-interference-plus-noise-ratio (SINR) and rate coverage in a multi-tier hybrid {sub-6GHz-mmWave-THz} network, where each tier has a particular base station (BS) density, transmit power, bandwidth, number of BS antennas, and cell-association bias factor. The proposed framework incorporates the effects of sub-6 GHz, mmWave and THz channel characteristics, BS beamforming gain, and blockages.
We investigate the downlink (DL) and uplink (UL) decoupled cell-association strategy and characterise the per-tier cell-association probability. Based on that,
we analytically derive the  SINR and rate coverage probabilities for both DL and UL transmissions. The analytical results are validated via extensive Monte Carlo simulations.  
Numerical results demonstrate the superiority of the DL and UL decoupled cell-association strategy in terms of SINR and rate coverage over its coupled counterpart. 

\end{abstract}
\IEEEpeerreviewmaketitle
\section{Introduction}
The use of millimetre wave (mmWave) bands has been regarded as a key driver of network capacity gains for the fifth-generation (5G) cellular networks \cite{shokri2015millimeter}. However, future mobile traffic will grow exponentially due to the emerging  data-hungry applications such as holographic telepresence, virtual reality, and autonomous vehicles \cite{wang2023road}.  In this regard, the launch of the sixth generation (6G) cellular networks is inevitable. 
Recently, terahertz (THz) mobile communication has gained momentum rapidly and  been envisioned as a promising solution to meet the extremely high  data rate requirements of 6G \cite{rappaport2019wireless}. Compared with sub-6 GHz and mmWave frequency bands, the THz band (0.1-10 THz) is susceptible to unique propagation challenges such as ultra-high free-space path loss and molecular absorption loss caused by water vapours or oxygen molecules \cite{han2022terahertz}. Due to limited coverage, standalone THz networks may not suffice to provide ubiquitous and reliable wireless transmissions. This brings the need to evolve towards a hybrid {sub-6GHz-mmWave-THz} ecosystem to support more reliable high-rate communications. 
Heterogeneity is a key feature of sub-6 GHz and mmWave cellular networks \cite{rangan2014millimeter}, where high-power base stations (BSs) with low densities coexist with denser low-power small-cell BSs. To compensate for the high free-space path loss at mmWave and THz frequencies, large antenna arrays are deployed at the BSs to provide adaptive directivity while avoiding inter-cell interference. The deployment of heterogeneous antenna arrays at different types of BSs increases the heterogeneity of millimeter wave and terahertz networks \cite{deng2018millimeter}. This motivates the modelling of a multi-tier heterogeneous network.

In this paper, we present a multi-tier hybrid {sub-6GHz-mmWave-THz} network model where the band-specific channel propagation characteristics are explicitly modelled and each mmWave/THz BS is equipped with a large antenna array to compensate for the propagation loss.  Different from the existing work investigating mmWave and THz networks, we focus on the impact of cell association on the signal-to-interference-plus-noise-ratio (SINR) and rate coverage, and highlight the advantages of applying the downlink (DL) and uplink (UL) decoupled cell-association strategy in comparison to its coupled counterpart.

\subsection{Related Works}
Standalone mmWave or THz networks have been extensively investigated using the tools from stochastic geometry \cite{bai2014coverage, turgut2017coverage, Chen2021Optimal, chen2021coverage, wu2020interference}. Due to the higher penetration loss through blockages at mmWave frequencies than at sub-6 GHz frequencies, line-of-sight (LOS)
and non-line-of-sight (NLOS) links need to be appropriately modelled.
In \cite{bai2014coverage}, the authors adopted a sectored model to approximate the antenna array gain and a LOS ball model, where the LOS region is assumed to be a ball with a fixed radius centred at the receiver of interest, to approximate the effect of blockages in a single-tier mmWave cellular network. The similar analytical methods were applied to heterogeneous mmWave cellular networks in \cite{turgut2017coverage}. In \cite{Chen2021Optimal}, the authors extended the analysis to a 3D scenario through considering BS heights and modelling the blockages as cylinders. Taking into account propagation characteristics of the THz band, \cite{chen2021coverage} and \cite{wu2020interference} analysed the coverage performance for outdoor and indoor THz networks, respectively. Modelling transmitters  and receivers as blockages, the authors in \cite{chen2021coverage} showed that excessive THz nodes can adversely affect the coverage performance. Considering the blockage effects of interior walls and random human bodies, it was shown in \cite{wu2020interference} that there exists an optimal THz BS density that maximises indoor coverage.

Lately, a handful of works studied the deployment of THz networks over the existing sub-6 GHz or mmWave networks \cite{sayehvand2020interference, humadi2022user, sopin2022user, hossan2021mobility}. In \cite{sayehvand2020interference}, the authors characterised the DL interference and coverage probability a coexisting THz and sub-6 GHz network; the analysis revealed that biased received signal power association can achieve a better coverage performance than the conventional reference signal received power association. \cite{humadi2022user} addressed the interference alleviation problem using a user-centric network design in an ultra-dense {sub-6GHz-mmWave-THz} network. \cite{sopin2022user} compared different user association and multi-connectivity methods in a two-tier mmWave-THz network, considering the micromobility of user equipment (UE). In \cite{hossan2021mobility}, the authors investigated UE mobility in a two-tier sub-6GHz-THz network and characterised the handoff probability. To date, a general multi-tier hybrid {sub-6GHz-mmWave-THz} network framework is still missing. Moreover, none of the research works evaluated the coverage and rate performance for UL transmissions.

The demand of UL communications has increased significantly along with the
evolution of social networking and mobile edge computing. In this context,
the DL and UL decoupled cell association plays an important role in improving network performance with regard to SINR and rate, especially for UL transmissions, in heterogeneous networks (HetNets) \cite{boccardi2016decouple}. Different from the coupled access where each UE is connected to the same BS during DL and UL communications, the decoupled cell-association strategy allows separate cell-association decisions in the DL and UL. The existing work exploring DL and UL decoupled access mainly focuses on co-channel HetNets \cite{lahad2020joint, dai2020decoupled}. In \cite{Elshaer2016Downlink}, the authors studied DL and UL decoupled access in a two-tier sub-6GHz-mmWave HetNet. However, the analysis does not apply to hybrid {sub-6GHz-mmWave-THz} networks.

To the best of our knowledge, the DL and UL performance analysis for a multi-tier hybrid {sub-6GHz-mmWave-THz} network under a DL and UL decoupled cell-association strategy
 has not been investigated yet, which motivates this work.

\subsection{Contributions}
The main goal of this paper is to analyse the SINR and rate coverage performance of a multi-tier hybrid {sub-6GHz-mmWave-THz} networks for both DL and UL transmissions. 
In particular, we investigate the DL and UL decoupled cell-association strategy and provide novel insights into cell-association bias design.
The main contributions of this paper are summarised as follows:
\begin{itemize}
\item We develop a novel stochastic geometry-based mathematical framework for the performance analysis of a general multi-tier hybrid {sub-6GHz-mmWave-THz} network,
capturing the effects of sub-6 GHz, mmWave and THz channel propagation characteristics, large-scale antenna array gain, and LOS probability due to blockages in the environment.

\item Based on the analytical framework, we derive the DL and UL per-tier cell-association probabilities under a flexible DL and UL decoupled cell-association strategy, where separate bias factors are used for DL and UL cell associations. Subsequently, we discuss the impact of THz BS density and bias on cell association.

\item Utilising the cell-association probabilities established, we newly derive tractable expressions of the SINR and rate coverage probabilities in the whole network or a certain sub-6 GHz/mmWave/THz tier for both DL and UL communications. We perform extensive numerical simulations to validate and show the effects of THz BS density, number of antennas, molecular absorption coefficient, and bias on SINR and rate coverage.

\item We quantitatively demonstrate the benefits of applying the DL and UL decoupled cell-association strategy in terms of providing good SINR and rate coverage for both DL and UL communications. 
\end{itemize}

\subsection{Paper Organisation}

The remainder of this paper is structured as follows. Section II describes the system model. In the section III, the per-tier cell-association probability is characterised. The expressions of DL and UL SINR coverage probabilities are derived in Section IV. In Section V, we extend our analysis to rate coverage probability. The numerical results are presented in Section VI. Finally, the conclusions are drawn in Section VII.

%\begin{figure}[t]
%\centerline{\includegraphics[scale=0.21]{Fig1.png}}
%\caption{An illustration of a HetNet with hybrid mmWave and THz.}
%\label{fig1}
%\end{figure}

\section{System Model}
\label{sec:system_model}
In this section, we introduce the system model of a general $K$-tier hybrid {sub-6GHz-mmWave-THz} network. We present the spatial network deployment, blockage, BS beamforming, and band-specific propagation model. The notations used in this paper are listed in Table \ref{tab1}.

\begin{table}[t]
\caption{Summary of Notations}
\begin{center}
\scalebox{1}{
\begin{tabular}{|c|c|}
\hline   
\textbf{Notation}&\textbf{Meaning}\\
\hline  
\hline
$\mathcal{K}$, ${\mathcal{S}}$, $\mathcal{M}$, $\mathcal{T}$ & \tabincell{c}{Sets of indices of the whole $K$-tier network, \\ sub-6 GHz, mmWave, and THz tiers, \\ respectively }  \\
\hline
$\Phi_{k}$, $\Phi_{\mathrm{b}}$, $\Phi_{\mathrm{U}}$ & \tabincell{c}{HPPPs modelling the locations of BSs in\\ the $k^{\mathrm{th}}$ tier, blockages and UE, respectively}  \\
\hline
$\lambda_{k}$, $\lambda_{\mathrm{b}}$, 
$\lambda_{\mathrm{U}}$ & \tabincell{c}{Densities of BSs in the $k^{\mathrm{th}}$ tier, \\ blockages and UE, respectively}  \\
\hline
${U_{0}}$ & The typical UE \\
\hline
${B_{k}^{\mathrm{UL}}}$ & \tabincell{c}{The typical BS in the $k^{\mathrm{th}}$ tier} \\
\hline
$L$, $W$ & \tabincell{c}{Mean values of blockage length and \\ blockage width, respectively} \\
\hline 
$P_{\mathrm{LOS}}(x)$ & LOS probability at distance $x$  \\
\hline
$N_{k}$ & Number of antennas per BS in the $k^{\mathrm{th}}$ tier \\
\hline
$G_{k}^{\mathrm{max}}$, $G_{k}^\mathrm{min}$  &  \tabincell{c}{Main-lobe and side-lobe beamforming gains \\ for the $k^{\mathrm{th}}$ tier, respectively}  \\
\hline
$P_{\mathrm{G},k}^{\mathrm{max}}$, $P_{\mathrm{G},k}^{\mathrm{min}}$ & \tabincell{c}{Probabilities of main-lobe and side-lobe gains \\ for the $k^{\mathrm{th}}$ tier, respectively} \\
\hline
${l^{\mathrm{S}}_{s}(x)}$, $l^{\mathrm{M}}_{m}(x)$, $l^{\mathrm{T}}_{t}(x)$ & \tabincell{c}{Propagation losses for the $s^{\mathrm{th}}$ sub-6 GHz tier, \\ the $m^{\mathrm{th}}$ mmWave tier, and the $t^{\mathrm{th}}$ THz tier \\ at distance $x$, respectively} \\
\hline
${\alpha_{s}}$, $\alpha_{m}$, $\alpha_{t}$ &\tabincell{c}{Path loss exponents of the $s^{\mathrm{th}}$ sub-6 GHz \\ tier,  the $m^{\mathrm{th}}$ mmWave tier, and the $t^{\mathrm{th}}$ THz \\ tier, respectively} \\
\hline 
${h_{s}}$, $h_{m}$ & \tabincell{c}{Small-scale fading power gain for the \\ $s^{\mathrm{th}}$ sub-6 GHz and $m^{\mathrm{th}}$ mmWave tiers} \\
\hline
$\gamma_{m}$ & \tabincell{c}{Shape parameter of the small-scale fading \\ power gain in the $m^{\mathrm{th}}$ mmWave tier} \\
\hline
$f_{\mathrm{M}}$, $f_{\mathrm{T}}$ & \tabincell{c}{Transmission frequencies for mmWave and \\ THz tiers, respectively} \\
\hline
$K_a$ & \tabincell{c}{Molecular absorption coefficient for THz tiers} \\
\hline
${R_{s}}$ & \tabincell{c}{Distance from the typical UE to its nearest \\ BS in the $s^{\mathrm{th}}$  sub-6 GHz tier} \\
\hline
$D_{m}$, $D_{t}$ & \tabincell{c}{Distances from the typical UE to its nearest \\ LOS BS in the $m^{\mathrm{th}}$   mmWave tier, \\ and the $t^{\mathrm{th}}$ THz tier,  respectively} \\
\hline
\tabincell{c}{${P_{s}^{q}}$, $P_{m}^{q}$, $P_{t}^{q}$,\\ $q\in\{\mathrm{DL, UL}\}$} & \tabincell{c}{Transmit powers in the $s^{\mathrm{th}}$ sub-6 GHz tier, \\ the $m^{\mathrm{th}}$ mmWave tier, and the $t^{\mathrm{th}}$ THz tier, \\ respectively}   \\
\hline
\tabincell{c}{${C_{s}^{q}}$, $C_{m}^{q}$, $C_{t}^{q}$, \\ $q\in\{\mathrm{DL, UL}\}$} & \tabincell{c}{Biased factors for the $s^{\mathrm{th}}$ sub-6 GHz tier, \\ the $m^{\mathrm{th}}$ mmWave tier, and the $t^{\mathrm{th}}$ THz tier, \\ respectively}  \\
\hline
\tabincell{c}{${\mathcal{A}_{s}^{q}}$, $\mathcal{A}_{m}^{q}$, $\mathcal{A}_{t}^{q}$,\\ $q\in\{\mathrm{DL, UL}\}$} & \tabincell{c}{Association probabilities that a typical UE is \\ connected to the $s^{\mathrm{th}}$ sub-6 GHz tier, the  \\ $m^{\mathrm{th}}$ mmWave tier, and the $t^{\mathrm{th}}$ THz tier, \\ respectively}   \\
\hline
\tabincell{c}{${X_{s}^{q}}$, $X_{m}^{q}$, $X_{t}^{q}$,\\ $q\in\{\mathrm{DL, UL}\}$} & \tabincell{c}{Distances from the typical UE to the serving  \\ BS in the $s^{\mathrm{th}}$ sub-6 GHz tier, the $m^{\mathrm{th}}$  \\ mmWave tier, and the $t^{\mathrm{th}}$ THz tier, \\ respectively} \\
\hline
$\tau, \rho$ & SINR and rate thresholds, respectively \\
\hline
${\delta_{s}^2}$, $\delta_{m}^2$, $\delta_{t}^2$ & \tabincell{c}{Noise powers  in the $s^{\mathrm{th}}$ sub-6 GHz tier,\\ the $m^{\mathrm{th}}$ mmWave tier, and the $t^{\mathrm{th}}$ THz tier, \\ respectively}   \\
\hline
$B_{\mathrm{W},k}$ & Bandwidth of the $k^{\mathrm{th}}$ tier \\
\hline
\tabincell{c}{$Z_{k}^{q}$,  $q\in\{\mathrm{DL, UL}\}$} & Average traffic load in the $k^{\mathrm{th}}$ tier \\
\hline
\end{tabular}}
\label{tab1}
\end{center}
\end{table}

\subsection{Network Model}
\subsubsection{{Hybrid sub-6GHz-mmWave-THz networks}}
%PPP of BS with total $K$-tier, $M$-tier mmWave and $T$-tier THz ($K=M+T$)...
In this paper, we consider a $K$-tier {hybrid sub-6GHz-mmWave-THz network}, where the locations of BSs in the $k^{\mathrm{th}}$ tier are modelled following a homogeneous Poisson point process (HPPP) $\Phi_{k}$ with density $\lambda_{k}$ on the two-dimensional (2D) plane. More specifically, { the sub-6GHz-mmWave-THz network consists of $S$ tiers of sub-6 GHz BSs}, $M$ tiers of mmWave BSs and $T$ tiers of THz BSs, where $K=S+M+T$. {The set of indices of sub-6 GHz tiers is denoted by $\mathcal{S} = \{1, 2, \dots, S\}$, that of mmWave tiers is denoted by $\mathcal{M} = \{S+1, S+2, \dots, S+M\}$ and that of THz tiers is denoted by $\mathcal{T} = \{S+M+1, S+M+2, \dots, S+M+T\}$.} Accordingly, the set of indices of all the tiers is denoted by {$\mathcal{K}=\{\mathcal{S}, \mathcal{M}, \mathcal{T}\}$.}
{The BSs in the $s^{\mathrm{th}}$ sub-6 GHz tier, where $s\in \mathcal{S}$, the $m^{\mathrm{th}}$ mmWave tier, where $m\in \mathcal{M}$, and the $t^{\mathrm{th}}$ THz tier, where $t\in \mathcal{T}$, are distributed following three independent HPPPs $\Phi_{s}$, $\Phi_{m}$ and $\Phi_{t}$ with densities $\lambda_{s}$, $\lambda_{m}$ and $\lambda_{t}$, respectively.} The locations of UE follow an independent HPPP $\Phi_{\mathrm{U}}$ with density $\lambda_{\mathrm{U}}$. Without loss of generality, we evaluate the performance of the typical UE located at the origin, {which is denoted by $U_0$. 
The BS that is serving the typical UE during UL transmissions is referred to as the typical BS. When the typical UE is associated with the $k^{\mathrm{th}}$ tier in the UL, the typical BS is denoted by $B_{k}^{\mathrm{UL}}$.}
We assume that $\lambda_{\mathrm{U}}\gg \lambda_{k}, \forall k\in \mathcal{K}$, so that each BS may serve multiple UE \cite{bai2014coverage, wang2023performance, ding2017uplink}. Intra-cell interference is eliminated by using orthogonal time/frequency resource partitioning.

\subsubsection{Blockage}
%Random size rectangle, PPP distribution, LOS probaility...
We focus on outdoor networks and consider buildings as the main blockages. Utilising random shape theory, we model the building as randomly sized rectangles \cite{bai2012using}. The centres of the blockages are distributed following an HPPP $\Phi_{\mathrm{b}}$ with density $\lambda_{\mathrm{b}}$. The length of the blockages $l_{\mathrm{b}}$ follows an arbitrary probability density function (PDF) $f_{L}(l_{\mathrm{b}})$ with mean $L$, and the width of the blockages $w_{\mathrm{b}}$ follows another arbitrary PDF $f_{W}(w_{\mathrm{b}})$ with mean $W$. Accordingly, the LOS probability of the transmission link from a BS to the typical UE is given by \cite{bai2014analysis}:
\begin{align}
P_{\mathrm{LOS}}(d) = e^{-(\zeta d + p)},
\end{align}
where $\zeta = \frac{2\lambda_{\mathrm{b}}(L+W)}{\pi}$, $p=\lambda_{\mathrm{b}}LW$, and $d$ is the distance between the BS and the typical UE.

\subsection{Beamforming Model}
Large antenna arrays are deployed on the mmWave and THz BSs to perform directional beamforming. On the other hand, the sub-6 GHz BSs are assumed to be equipped with a single antenna \cite{Elshaer2016Downlink}.
It particular, each BS in the $k^{\mathrm{th}}$ tier is equipped with a uniform linear array with $N_{k}$ antenna elements. We have that $N_{k}\gg 1$ if $k\in \{\mathcal{M}, \mathcal{T}\}$ and $N_{k}= 1$ if $k\in \{\mathcal{S}\}$. The inter-element spacing is assumed to be half of the wavelength. Each UE is equipped with a single receiving antenna. Due to the excessive power consumption of RF chain components at mmWave/THz frequencies, we adopt analog beamforming to provide directional beams. 

We assume that each BS can align its beam to its serving UE to achieve the maximum beamforming gain. For the $k^{\mathrm{th}}$ tier, the actual BS antenna radiation pattern is computed by the Fej\'{e}r kernel function in linear scale as follows
\begin{align}
\label{eq22}
G_{k}(N_{k},\phi_{B_{b,k}}, \phi_{S_{b,k}}) =\frac{\mathrm{sin}^{2}\left(\frac{\pi N_{k}}{2}\left(\mathrm{cos}\phi_{B_{b,k}}-\mathrm{cos}\phi_{S_{b,k}}\right)\right)}{N_{k}\mathrm{sin}^{2}\left(\frac{\pi}{2}\left(\mathrm{cos}\phi_{B_{b,k}}-\mathrm{cos}\phi_{S_{b,k}}\right)\right)},
\end{align}
where $B_{b,k}$ denotes BS $b$ in the $k^{\mathrm{th}}$ tier, $\phi_{B_{b,k}}$ is the azimuth angle between $B_{b,k}$ and the typical UE, and $\phi_{S_{b,k}}$ is the azimuth angle between $B_{b,k}$ and its served UE. If $B_{b,k}$ is the serving BS of the typical UE, we have $\phi_{B_{b, k}}=\phi_{S_{b,k}}$ and $G_{k}(N_{k}) = N_k$; if $B_{b,k}$ is an interfering BS, $\phi_{{D}_{b, k}}=\frac{1}{2}\left(\mathrm{cos}\phi_{B_{b,k}}-\mathrm{cos}\phi_{S_{b,k}}\right)$ is uniformly distributed over $[-0.5,0.5]$ \cite{Lee2016Randomly}, and (\ref{eq22}) is rewritten as follows
\begin{align}
\label{eq3}
G_{k}(N_{k}, \phi_{{D}_{b,k}}) =  \frac{\mathrm{sin}^{2}\left(\pi N_{k}\phi_{{D}_{b,k}}\right)}{N_{k}\mathrm{sin}^{2}\left(\pi \phi_{{D}_{b,k}}\right)}.
\end{align}
To enable tractable analysis,  we adopt a normalised flat-top BS antenna array radiation pattern proposed in \cite{deng2018millimeter} to approximate (\ref{eq3}), which is expressed as
\begin{align}
\label{eq:flat_antenna}
G_{\mathrm{flat},k}(N_{k}, \phi_{{D}_{b,k}})=
\left\{
\begin{array}{ll}
G_{k}^{\mathrm{max}}, & { |\phi_{{D}_{b,k}}| \leq \phi_{\mathrm{3dB},k},}
\\
G_{k}^{\mathrm{min}}, &  {\mathrm{otherwise},} 
\end{array} \right.
\end{align}
where $G_{k}^{\mathrm{max}}=N_{k}$ is the main-lobe beamforming gain, $G_{k}^{\mathrm{min}}=\frac{1-2 \phi_{\mathrm{3dB},k}G_{k}^{\mathrm{max}}}{1-2\phi_{\mathrm{3dB},k}}$ is the side-lobe  beamforming gain, and $\phi_{\mathrm{3dB},k}$ is the half-power beamwidth (HPBW) of beamforming gain, which is calculated by  $G_{k}(N_{k},\phi_{\mathrm{3dB},k})=\frac{N_{k}}{2}$.

\subsection{Channel Model}
\label{sec:channel_model}
{In the considered hybrid sub-6GHz-mmWave-THz network, we assume dense deployments of mmWave BSs and
THz BSs. Given the low ratio of sub-6 GHz BSs to blockages, we assume that the sub-6 GHz transmission links are in NLOS conditions.}
Due to the high penetration loss of mmWave/THz transmission, the received signals from NLOS links are negligible compared to those from LOS links {\cite{yu2017coverage,Chen2021Optimal}}. Therefore, in this paper, we focus on LOS transmission links for the mmWave and THz tiers.
{The assumption of ignoring NLOS transmission links will be justified by Fig. \ref{fig:CP_NLOS_AS} in Section \ref{sec:numerical}.}

{
\subsubsection{Sub-6 GHz} 
In sub-6 GHz transmission, the channel model incorporates both large-scale path loss and small-scale fading. The total propagation loss from an NLOS sub-6 GHz BS in the $s^{\mathrm{th}}$ tier to the typical UE is expressed as
\begin{equation}
l^{\mathrm{S}}_{s}(d_s) = \beta_{0}d_{s}^{-\alpha_{s}}h_{s},
\end{equation}
where  $\beta_{0}$ is the path loss at the reference distance of 1 m, $d_{s}$ is the transmission distance, $\alpha_s$ is the path loss exponent in the $s^{\mathrm{th}}$ tier, and $h_{s}$ is the power gain of small-scale fading in the $s^{\mathrm{th}}$ tier. The small-scale fading in sub-6 GHz tiers is modelled as a Rayleigh distribution, i.e., $h_{s}\sim \mathrm{Exp}(1)$.
}

\subsubsection{mmWave}
The channel model in mmWave transmission also consists of large-scale path loss and small-scale fading.
The total propagation loss from an LOS mmWave BS in the $m^{\mathrm{th}}$ tier to the typical UE is given by
\begin{equation}
l^{\mathrm{M}}_{m}(d_m) = \left(\frac{c}{4\pi f_{\mathrm{M}}}\right)^{2}d_{m}^{-\alpha_{m}}h_{m},
\end{equation}
where $c$ is the speed of light, $f_{\mathrm{M}}$ is the mmWave transmission frequency, $d_{m}$ is the transmission distance, $\alpha_m$ is the path loss exponent in the $m^{\mathrm{th}}$ tier, and $h_{m}$ is the power gain of small-scale fading in the $m^{\mathrm{th}}$ tier. In this paper, we model the small-scale fading in mmWave tiers as Nakagami-$m$ distribution with $h_{m}\sim\Gamma(\gamma_m,\frac{1}{\gamma_m})$ \cite{bai2014coverage}, where $\gamma_m$ is the shape parameter of small-scale fading power gain in the $m^{\mathrm{th}}$ tier. 

\subsubsection{THz}
%carrier frequency, absorption coefficient
For THz propagation, we need to further consider the effect of molecular absorption.
The total propagation loss from an LOS THz BS in the $t^{\mathrm{th}}$ tier to the typical UE is given by
\begin{equation}
l^{\mathrm{T}}_{t}(d_t) = \left(\frac{c}{4\pi f_{\mathrm{T}}}\right)^{2}d_{t}^{-\alpha_{t}}e^{-K_ad_t},
\end{equation}
where $f_{\mathrm{T}}$ is the THz transmission frequency, $d_{t}$ is the transmission distance, $\alpha_t$ is the path loss exponent in the $t^{\mathrm{th}}$ tier, and $K_a$ is the molecular absorption coefficient. Note that the small-scale fading is negligible at THz bands.
The molecular absorption coefficient is intricately influenced by the ambient environmental conditions, including factors such as atmospheric composition, humidity levels, and the specific frequency of transmission \cite{Jornet2011Channel}.  For simplicity, we will not elaborate on the modelling of the molecular absorption coefficient. The analytical results of this paper are applicable to any value of $K_a$.

\section{Downlink and Uplink Decoupled Cell Association}

In this section, we first characterise the PDF of the distance from the typical UE to its nearest LOS BS in the $k^{\mathrm{th}}$ tier. Then we derive the cell-association probability for DL and UL transmissions, respectively.
\begin{lemma}
\label{lemma:fRk}
Denoting by $D_{k}$ the distance from the typical UE to its nearest LOS BS in the $k^{\mathrm{th}}$ tier, the PDF of $D_{k}$ is given by  $f_{D_{k}}(x) = 2\pi\lambda_{k}x \, \mathrm{exp}\!\!\left(\frac{2\pi\lambda_{k}e^{ -(\zeta x + p)}\left(1-e^{\zeta x}+\zeta x\right)}{\zeta^2} -\zeta x -p \right)$.
\end{lemma}
\begin{IEEEproof}
The cumulative distribution function (CDF) of $D_{k}$ is computed by 
\begin{align}
 F_{D_{k}}(x)&=1-\mathbb{P}(D_{k}>x) \nonumber \\
 &\overset{(a)}{=}1- \mathrm{exp}\left(-2\pi \lambda_{k}\int_{0}^{x}e^{-(\zeta D_{k} + p)}D_{k}\mathrm{d}D_{k}\right) \nonumber \\
 &=1- \mathrm{exp}\left(\frac{2\pi\lambda_{k}e^{ -(\zeta x + p)}\left(1-e^{\zeta x}+\zeta x\right)}{\zeta^2}\right),
\end{align}
where (a) is obtained using the void probability of HPPP with LOS probability. More specifically, for an HPPP with intensity $\lambda$, the void probability of finding no points in a region with radius $x$ is given by $P_{\mathrm{void}}\!\!=\!\mathrm{exp}\left(\!-\!2\pi \lambda\int_{0}^{x}r\text{d}r\right)\!\!=\!\mathrm{exp}(\!-\pi\lambda x^2)$. Then the PDF of $D_{k}$ is computed by
$f_{D_{k}}(x) = \frac{\mathrm{d}F_{D_{k}}(x)}{\mathrm{d}\, x}$.
\end{IEEEproof}

\begin{figure}[!t]
 \centering
 \subfloat[]{\includegraphics[width=3in]{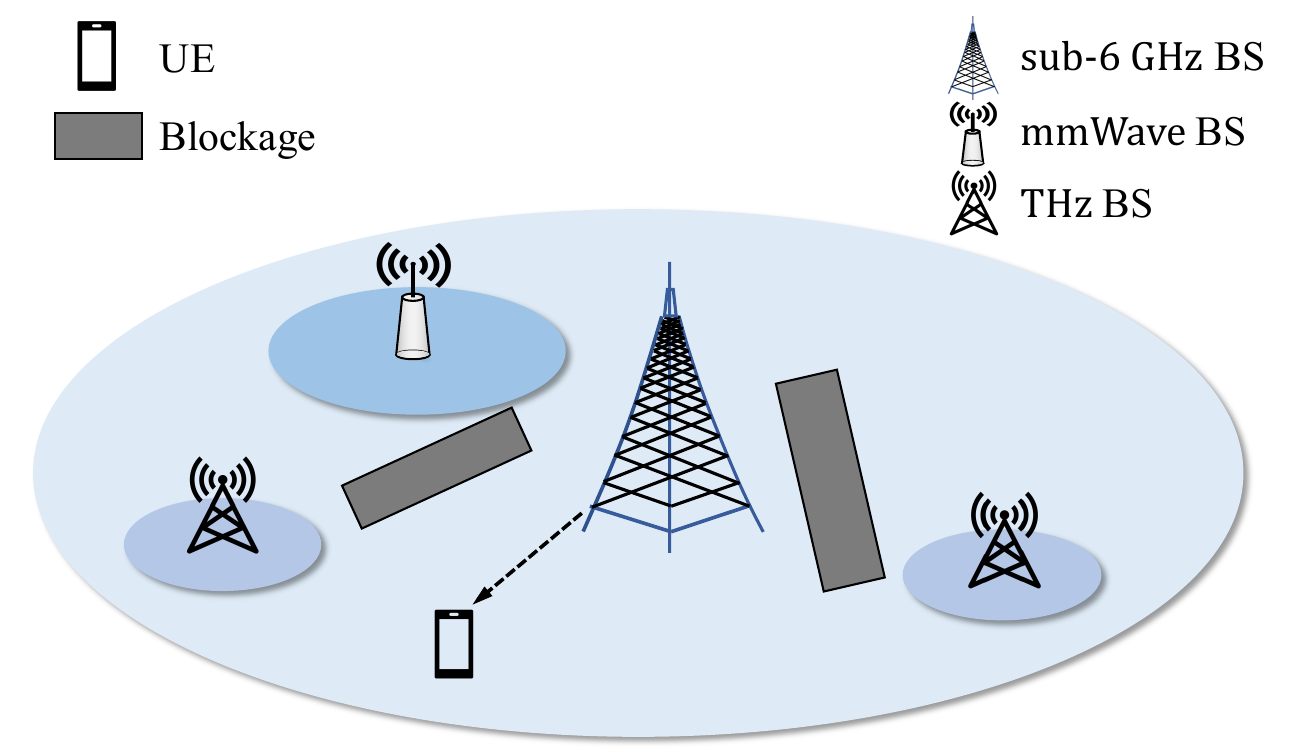}} \\
 \subfloat[]{\includegraphics[width=3in]{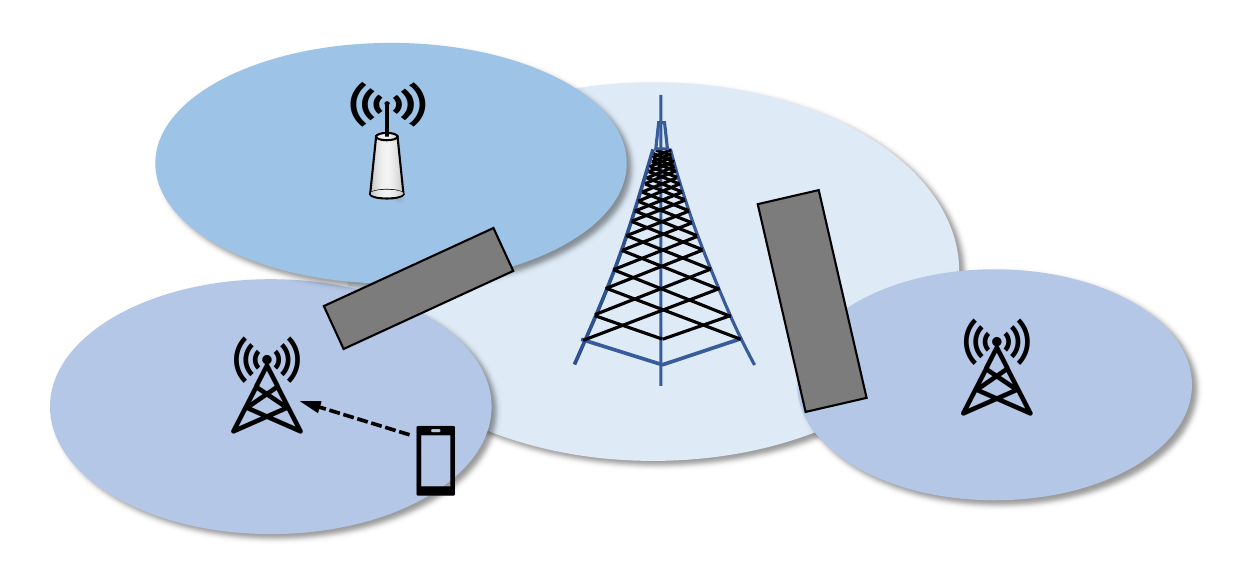}}
 \caption{{Illustration of DL and UL decoupled cell association in a 3-tier hybrid sub-6GHz-mmWave-THz network. (a) DL. (b) UL.}}
\label{fig:association}
\end{figure}

In this paper, we consider DL and UL decoupled cell association. 
Due to the difference in DL and UL transmission powers, each cell may have different DL and UL coverage areas. As such, each UE may be associated with different BSs during DL and UL transmissions, as shown in Fig. \ref{fig:association}. To this end, we investigate a flexible cell-association strategy where each UE connects to the BS that provides the strongest average
biased received signal for both DL and UL communications. 

{The average biased received power {in the DL} at the typical UE from the nearest sub-6 GHz BS in the $s^{\mathrm{th}}$ tier {(in the UL at the nearest sub-6 GHz BS in the $s^{\mathrm{th}}$ tier from the typical UE)}  is given by
\begin{equation}
\label{eq:sub6_RP}
P^{\mathrm{S}, q}_{r,s}(R_{s}) = P_{s}^{q}l^{\mathrm{S}}_{s}(R_{s})C_{s}^{q},
\end{equation}
where $q\in\{\mathrm{DL, UL}\}$, $R_{s}$ is the distance from the typical UE to the nearest sub-6 GHz BS in the $s^{\mathrm{th}}$ tier, $P_{s}^{\mathrm{DL}}$ is the DL transmit power of BSs in the $s^{\mathrm{th}}$ tier, $P_{s}^{\mathrm{UL}}$ is the UL transmit power of UE connected to the $s^{\mathrm{th}}$ tier, and $C_{s}^{\mathrm{DL}}$ and $C_{s}^{\mathrm{UL}}$ are the DL and UL cell-association bias factors of the $s^{\mathrm{th}}$ tier, respectively. If $C_{s}^{\mathrm{DL}} \left(C_{s}^{\mathrm{UL}}\right)>1$, more UE will be offloaded to the $s^{\mathrm{th}}$ tier in the DL (UL) transmission. }
Similarly, the average biased received power {in the DL} at the typical UE from the nearest LOS mmWave BS in the $m^{\mathrm{th}}$ tier {(in the UL at the nearest LOS mmWave BS in the $m^{\mathrm{th}}$ tier from the typical UE)} is given by
\begin{equation}
\label{eq:mmWave_RP}
P^{\mathrm{M}, q}_{r,m}(D_{m}) = P_{m}^{q}G_{m}^{\mathrm{max}}l^{\mathrm{M}}_{m}(D_{m})C_{m}^{q},
\end{equation}
where $q\in\{\mathrm{DL, UL}\}$. The average biased received power {in the DL} at the typical UE from the nearest LOS THz BS in the $t^{\mathrm{th}}$ tier {(in the UL at the the nearest LOS THz BS in the $t^{\mathrm{th}}$ tier from the typical UE)} is given by
\begin{equation}
\label{eq:Thz_RP}
P^{\mathrm{T}, q}_{r,t}(D_{t}) = P_{t}^{q}G_{t}^{\mathrm{max}}l^{\mathrm{T}}_{t}(D_{t})C_{t}^{q},
\end{equation}
where $q\in\{\mathrm{DL, UL}\}$.

{
\subsection{Sub-6 GHz Cell Association}
In the Lemma below, we determine the probability that the typical UE associates with a sub-6 GHz BS in the $s^{\mathrm{th}}$ tier.
\begin{lemma}
	\label{lemma_sub6_AP}
	The probability that the typical UE is associated with the $s^{\mathrm{th}}$ sub-6 GHz tier for the DL and UL is given by
	\begin{align}
	&\mathcal{A}^{q}_{s} = \int_{0}^{\infty} f_{R_{s}}(x)  \prod_{g\in \mathcal{S},g\neq s} \mathrm{exp}\left( -\pi\lambda_{g} \left(\varrho_{s,g}^{q}(x)\right)^2 \right) \times  \nonumber\\
	&\prod_{i\in \mathcal{M}} \! \mathrm{exp}\!\left( \!\! \frac{2\pi\lambda_{i} e^{ -(\zeta \varepsilon_{s,i}^{q}(x) + p)}\!\!\left(\!1 \!-\! e^{\zeta \varepsilon_{s,i}^{q}(x)} \!+\! \zeta \varepsilon_{s,i}^{q}(x)\!\right)}{\zeta^2} \!\! \right) \! \times  \nonumber\\
    &\prod_{j\in \mathcal{T}} \! \mathrm{exp}\!\left( \!\! \frac{2\pi\lambda_{j} e^{ -(\zeta \vartheta_{s,j}^{q}(x) + p)}\!\!\left(\!1 \!-\! e^{\zeta \vartheta_{s,j}^{q}(x) } \!+\! \zeta \vartheta_{s,j}^{q}(x)\!\right)}{\zeta^2} \!\! \right) \mathrm{d}x,
	\end{align}
	where $q\in\{\mathrm{DL, UL}\}$, $\varrho_{s,g}^{q}(x)=\left(\frac{C_g^{q}P_g^{q}}{C_s^{q}P_s^{q}}\right)^{\frac{1}{\alpha_{g}}}x^{\frac{\alpha_{s}}{\alpha_{g}}}$, $\varepsilon_{s,i}^{q}(x)=\left(\frac{C_i^{q}P_i^{q}N_i}{C_s^{q}P_s^{q}\beta_{0}}\left(\frac{c}{4\pi f_{\mathrm{M}}}\right)^{2}\right)^{\frac{1}{\alpha_{i}}}x^{\frac{\alpha_{s}}{\alpha_{i}}}$, and $\vartheta_{s,j}^{q}(x)=\frac{\alpha_{j}}{K_a}W_{\mathrm{L}}\left(\Lambda_{s,j}^{\mathrm{S},q}x^{\frac{\alpha_{s}}{\alpha_{j}}}\right)$, in which $W_{\mathrm{L}}(\cdot)$ is the Lambert W function with $x=W_{\mathrm{L}}(x)e^{W_{\mathrm{L}}(x)}$ and
	$\Lambda_{s,j}^{\mathrm{S},q}=\frac{K_a}{\alpha_{j}}\left(\frac{C_j^{q}P_j^{q}N_j}{C_s^{q}P_s^{q}\beta_{0}}\left(\frac{c}{4\pi f_{\mathrm{T}}}\right)^{2} \right)^{\frac{1}{\alpha_j}}$. Besides, $f_{R_{s}}(x)=2\pi\lambda_s xe^{-\pi\lambda_s x^2}$ is the PDF of the distance from the typical UE to its nearest BS in the $s^{\mathrm{th}}$ tier.
\end{lemma}
\begin{IEEEproof}
	See Appendix \ref{App:sub6_AP}.
\end{IEEEproof}

Letting $X_{s}^{q}$, where $q\in\{\mathrm{DL, UL}\}$, be the distance from the serving BS to the typical UE, given that the typical UE is associated with a BS in the $s^{\mathrm{th}}$ sub-6 GHz tier, we characterise the PDF of $X_{s}^{q}$ in the following Lemma.
\begin{lemma}
	\label{lemma_sub6_PDF}
	The PDF of the distance between the serving BS in the $s^{\mathrm{th}}$ sub-6 GHz tier and the typical UE  for the DL and UL is given by 
	\begin{align}
	&f_{X_{s}^{q}}(x) = \frac{f_{R_{s}}(x)}{\mathcal{A}^{q}_{s}}  \prod_{g\in \mathcal{S},g\neq s} \mathrm{exp}\left( -\pi\lambda_{g} \left(\varrho_{s,g}^{q}(x)\right)^2 \right) \times  \nonumber\\
	&\prod_{i\in \mathcal{M}} \! \mathrm{exp}\!\left( \!\! \frac{2\pi\lambda_{i} e^{ -(\zeta \varepsilon_{s,i}^{q}(x) + p)}\!\!\left(\!1 \!-\! e^{\zeta \varepsilon_{s,i}^{q}(x)} \!+\! \zeta \varepsilon_{s,i}^{q}(x)\!\right)}{\zeta^2} \!\! \right) \! \times  \nonumber\\
    &\prod_{j\in \mathcal{T}} \! \mathrm{exp}\!\left( \!\! \frac{2\pi\lambda_{j} e^{ -(\zeta \vartheta_{s,j}^{q}(x) + p)}\!\!\left(\!1 \!-\! e^{\zeta \vartheta_{s,j}^{q}(x) } \!+\! \zeta \vartheta_{s,j}^{q}(x)\!\right)}{\zeta^2} \!\! \right).
	\end{align}
\end{lemma}

\begin{IEEEproof}
	See Appendix \ref{App:sub6_PDF}.
\end{IEEEproof}

}

\subsection{mmWave Cell Association}
In the following Lemma, the probability that the typical UE connects to an mmWave BS in the $m^{\mathrm{th}}$ tier is derived.
\begin{lemma}
\label{lemma_mmWave_AP}
The probability that the typical UE is associated with the $m^{\mathrm{th}}$ mmWave tier for the DL and UL is given by
\begin{align}
&\mathcal{A}^{q}_{m} = \int_{0}^{\infty} f_{D_{m}}(x) \prod_{g\in\mathcal{S}}\mathrm{exp}\left(-\pi\lambda_{g}\left(\chi_{s,g}^{q}(x)\right)^2\right) \times  \nonumber\\
&\prod_{i\in \mathcal{M},i\neq m} \!\!\!\!\!\!\!\mathrm{exp}\!\left(\!  \frac{2\pi\lambda_{i} e^{ -(\zeta \Omega_{m,i}^{q}(x) + p)}\left(1\!-\!e^{\zeta \Omega_{m,i}^{q}(x)} \!+\! \zeta \Omega_{m,i}^{q}(x)\right)}{\zeta^2} \!\!\right)  \times \nonumber\\
&\quad\prod_{j\in \mathcal{T}} \!\! \mathrm{exp}\left(\! \frac{2\pi\lambda_{j} e^{ -(\zeta \Psi_{m,j}^{q}(x) + p)}\!\!\left(\!1 \!\!-\! e^{\zeta \Psi_{m,j}^{q}(x)} \!\!+\! \zeta \Psi_{m,j}^{q}(x)\!\right)}{\zeta^2} \!\! \right) \!\!\mathrm{d}x,
\end{align}   
where $q\in\{\mathrm{DL, UL}\}$, $\chi_{m,g}^{q}(x)=\left(\frac{C_g^{q}P_g^{q}\beta_{0}}{C_m^{q}P_m^{q}N_m}\left(\frac{4\pi f_{\mathrm{M}}}{c}\right)^{2}\right)^{\frac{1}{\alpha_{g}}}x^{\frac{\alpha_{m}}{\alpha_{g}}}$, $\Omega_{m,i}^{q}(x)=\left(\frac{C_i^{q}P_i^{q}N_i}{C_m^{q}P_m^{q}N_m} \right)^{\frac{1}{\alpha_{i}}}x^{\frac{\alpha_{m}}{\alpha_{i}}}$, $\Psi_{m,j}^{q}(x)=\frac{\alpha_{j}}{K_a}W_{\mathrm{L}}\left(\Lambda_{m,j}^{\mathrm{M},q}x^{\frac{\alpha_{m}}{\alpha_{j}}}\right)$, and
$\Lambda_{m,j}^{\mathrm{M},q}=\frac{K_a}{\alpha_{j}}\left(\frac{f_{\mathrm{M}}^2P_j^{q}N_jC_j^{q}}{f_{\mathrm{T}}^2P_m^{q}N_mC_m^{q}} \right)^{\frac{1}{\alpha_j}}$.
\end{lemma}
\begin{IEEEproof}
	It follows similar proof as in Lemma \ref{lemma_sub6_AP}.
\end{IEEEproof}

Denote $X_{m}^{q}$, where $q\in\{\mathrm{DL, UL}\}$, as the distance from the serving BS to the typical UE, given that the typical UE is associated with a BS in the $m^{\mathrm{th}}$ mmWave tier. In the Lemma below, we characterise the PDF of $X_{m}^{q}$.
\begin{lemma}
\label{lemma_mmWave_PDF}
The PDF of the distance between the serving BS in the $m^{\mathrm{th}}$ mmWave tier and the typical UE  for the DL and UL is given by 
\begin{align}
&f_{X_{m}^{q}}(x) = \frac{f_{D_{m}}(x)}{\mathcal{A}^{q}_{m} } \prod_{g\in\mathcal{S}}\mathrm{exp}\left(-\pi\lambda_{g}\left(\chi_{s,g}^{q}(x)\right)^2\right) \times \nonumber\\ 
&\prod_{i\in \mathcal{M},i\neq m} \!\!\!\!\!\!\!\mathrm{exp}\!\!\left(\! \frac{2\pi\lambda_{i} e^{ -(\zeta \Omega_{m,i}^{q}(x) + p)}\!\left(\!1\!-\!e^{\zeta \Omega_{m,i}^{q}(x)} \!+\! \zeta \Omega_{m,i}^{q}(x) \! \right)}{\zeta^2} \!\!\right)  \!\!\times\! \nonumber\\
&\prod_{j\in \mathcal{T}} \! \mathrm{exp}\!\!\left(\! \frac{2\pi\lambda_{j} e^{ -(\zeta \Psi_{m,j}^{q}(x) + p)}\!\left(\!1 \!-\! e^{\zeta \Psi_{m,j}^{q}(x)} \!+\! \zeta \Psi_{m,j}^{q}(x) \! \right)}{\zeta^2} \!\! \right)\!\!.
\end{align}
\end{lemma}

\begin{IEEEproof}
	It follows similar proof as in Lemma \ref{lemma_sub6_PDF}.
\end{IEEEproof}

\subsection{THz Cell Association}
In the following Lemma, we present the probability that the typical UE links to a THz BS in the $t^{\mathrm{th}}$ tier.
\begin{lemma}
	\label{lemma_Thz_AP}
	The probability that the typical UE is associated with the $t^{\mathrm{th}}$ THz tier for the DL and UL is given by
	\begin{align}
	&\mathcal{A}^{q}_{t} = \int_{0}^{\infty} f_{D_{t}}(x) \prod_{g\in \mathcal{S}} \mathrm{exp}\left( -\pi\lambda_{g}\left(\Upsilon_{t,g}^{q}(x) \right)^2 \right) \times  \nonumber\\ 
	&\quad\prod_{i\in \mathcal{M}} \! \mathrm{exp}\!\!\left(\!\frac{2\pi\lambda_{i} e^{-(\zeta \Xi_{t,i}^{q}(x) + p)}\!\left(1-e^{\zeta \Xi_{t,i}^{q}(x)} + \zeta \Xi_{t,i}^{q}(x)\!\right)}{\zeta^2} \! \right) \!\times \nonumber\\
	&\prod_{j\in \mathcal{T},j\neq t} \!\!\!\!\!\! \mathrm{exp}\!\!\left(\! \frac{2\pi\lambda_{j} e^{ -(\zeta \Theta_{t,j}^{q}(x) + p)}\!\left(\!1 \!-\! e^{\zeta \Theta_{t,j}^{q}(x)} \!+\! \zeta \Theta_{t,j}^{q}(x)\!\right)}{\zeta^2} \! \right)  \!\mathrm{d}x,  
	\end{align}
where $q\in\{\mathrm{DL, UL}\}$, $\Upsilon_{t,g}^{q}(x)= \left(\frac{C_g^{q}P_g^{q}\beta_{0}}{C_t^{q}P_t^{q}N_t}\left(\frac{4\pi f_{\mathrm{T}}}{c}\right)^{2}\right)^{\frac{1}{\alpha_{g}}} e^{\frac{K_a}{\alpha_{g}}x}x^{\frac{\alpha_{t}}{\alpha_{g}}}$, $\Xi_{t,i}^{q}(x) \!=\! \left(\frac{C_i^{q}P_i^{q}N_if_{\mathrm{T}}^{2}}{C_t^{q}P_t^{q}N_tf_{\mathrm{M}}^{2}}\right)^{\frac{1}{\alpha_{i}}} \!\!\!\! e^{\frac{K_a}{\alpha_{i}}x}x^{\frac{\alpha_{t}}{\alpha_{i}}}$, and $\Theta_{t,j}^{q}(x)
	\!=\!
	\frac{\alpha_{j}}{K_a}W_{\mathrm{L}}(\Lambda_{t,j}^{\mathrm{T},q}e^{\frac{K_a}{\alpha_{j}}x}x^{\frac{\alpha_{t}}{\alpha_{j}}})$, in which $\Lambda_{t,j}^{\mathrm{T},q} \!=\! \frac{K_a}{\alpha_{j}} \left(\frac{C_j^{q}P_j^{q}N_j}{C_t^{q}P_t^{q}N_t}\right)^{\frac{1}{\alpha_{j}}}$.
\end{lemma}

\begin{IEEEproof}
	It follows similar proof as in Lemma \ref{lemma_sub6_AP}.
\end{IEEEproof}

  Let $X_{t}^{q}$, where $q\in\{\mathrm{DL, UL}\}$, be the distance from the serving BS to the typical UE, given that the typical UE is associated with a BS in the $t^{\mathrm{th}}$ THz tier. We characterise the PDF of $X_{t}^{q}$ in the Lemma below.

\begin{lemma}
	\label{lemma_Thz_PDF}
	The PDF of the distance between the serving BS in the $t^{\mathrm{th}}$ THz tier and the typical UE for the DL and UL is given by
	\begin{align}
	&f_{X_t^{q}}(x) = \frac{f_{D_{t}}(x)}{\mathcal{A}^{q}_{t} }  \prod_{g\in \mathcal{S}} \mathrm{exp}\left( -\pi\lambda_{g}\left(\Upsilon_{t,g}^{q}(x) \right)^2 \right)  \times \nonumber\\ 
	&\prod_{i\in \mathcal{M}} \! \mathrm{exp}\!\left(\!\frac{2\pi\lambda_{i} e^{-(\zeta \Xi_{t,i}^{q}(x) + p)}\!\left(1-e^{\zeta \Xi_{t,i}^{q}(x)} + \zeta \Xi_{t,i}^{q}(x)\!\right)}{\zeta^2} \! \right) \! \times \nonumber\\
	&\prod_{j\in \mathcal{T},j\neq t} \!\!\!\!\! \mathrm{exp}\!\left(\!  \frac{2\pi\lambda_{j} e^{ -(\zeta \Theta_{t,j}^{q}(x) + p)}\!\left(1 - e^{\zeta \Theta_{t,j}^{q}(x)} + \zeta \Theta_{t,j}^{q}(x)\!\right)}{\zeta^2} \! \right)\!.
	\end{align}	
\end{lemma}

\begin{IEEEproof}
	It follows similar proof as in Lemma \ref{lemma_sub6_PDF}.
\end{IEEEproof}

\section{SINR and Rate Coverage in Hybrid Sub-6GHz-mmWave-THz Networks}
%Muti-tier HetNet, Association strategy ...... 
In this section, we will elaborate on the performance analysis in terms of SINR and rate coverage probabilities for the hybrid sub-6GHz-mmWave-THz networks based on the stochastic geometry framework. 

\subsection{SINR Coverage Probability} 
The SINR coverage probability is used to quantify the reliability of wireless transmissions and is the basis of quantifying other performance metrics.
The DL (UL) SINR coverage probability is defined as the probability that the DL (UL) SINR of the typical UE is higher than a threshold $\tau$, which can be expressed as
{
\begin{align}
P^{q}_{\mathrm{cov}}(\tau)
\!\!= \!\! &\sum_{s\in \mathcal{S}}\!\mathcal{A}^{q}_{s}P^{q}_{\mathrm{cov}, s}(\tau)  +\!\!\!\! \sum_{m\in \mathcal{M}}\!\!\!\!\mathcal{A}^{q}_{m}P^{q}_{\mathrm{cov}, m}(\tau) +\!\! \sum_{t\in \mathcal{T}}\!\!\mathcal{A}^{q}_{t}P^{q}_{\mathrm{cov}, t}(\tau) \nonumber\\
= \!\!&\sum_{s\in \mathcal{S}}\mathcal{A}^{q}_{s}\int_{0}^{\infty}\mathbb{P}(\mathrm{SINR}^{q}_{s}(x)>\tau) f_{X_s^{q}}(x) \mathrm{d}x  \,\, +  \nonumber\\ 
&\!\!\sum_{m\in \mathcal{M}}\mathcal{A}^{q}_{m}\int_{0}^{\infty}\mathbb{P}(\mathrm{SINR}^{q}_{m}(x)>\tau) f_{X_m^{q}}(x) \mathrm{d}x  \,\, +  \nonumber\\ 
&\sum_{t\in \mathcal{T}}\mathcal{A}^{q}_{t}\int_{0}^{\infty}\mathbb{P}(\mathrm{SINR}^{q}_{t}(x)>\tau) f_{X_t^{q}}(x) \mathrm{d}x,
\end{align}
where $q\in\{\mathrm{DL, UL}\}$, $P_{\mathrm{cov}, s}^{q}(\tau)$, $P_{\mathrm{cov}, m}^{q}(\tau)$, and $P_{\mathrm{cov}, t}^{q}(\tau)$ represent the SINR coverage probabilities of a typical UE when it is associated with the $s^{\mathrm{th}}$ sub-6 GHz tier, the $m^{\mathrm{th}}$ mmWave tier, and the $t^{\mathrm{th}}$ THz tier, respectively. }

{
\subsubsection{Sub-6 GHz SINR Coverage Probability}
When the typical UE is associated with the $s^{\mathrm{th}}$ sub-6 GHz tier, the interference signals only come from the  sub-6 GHz tiers.
Let $R_{s}$ be $x$, the DL and UL received SINR can be expressed as
\begin{align}
\label{eq:sub6_SINR}
&\mathrm{SINR}^{q}_{s}(x) 
=\frac{P_{s}^{q}l^{\mathrm{S}}_{s}(x)}{I_{s}^{\mathrm{S},q} + \delta_{s}^2},
\end{align}
where $q\in\{\mathrm{DL, UL}\}$, $\delta_{s}^2$ is the additive white Gaussian noise (AWGN) power, $I_{s}^{\mathrm{S}, \mathrm{DL}}$ is the aggregated DL interference, which is given by
\begin{align}
\label{eq:sub6_interference_DL}
I_{s}^{\mathrm{S}, \mathrm{DL}}=\sum\limits_{j\in \mathcal{S}} \sum\limits_{i\in \Phi_{j}\backslash B_{s}^{\mathrm{DL}}}P^{\mathrm{DL}}_{j}l^{\mathrm{S}}_{j}(d_{i,j}),
\end{align}
where $B_{s}^{\mathrm{DL}}$ is the DL serving BS in the $s^{\mathrm{th}}$ sub-6 GHz tier, $\Phi_{j}$ is the set of BSs in the $j^{\mathrm{th}}$ tier, and $d_{i,j}$ is the distance from BS $i$ in the $j^{\mathrm{th}}$ tier to the typical UE, and $I_{s}^{\mathrm{S}, \mathrm{UL}}$ is the aggregated UL interference, which is given by
\begin{align}
\label{eq:sub6_interference_UL}
I_{s}^{\mathrm{S}, \mathrm{UL}}=\sum\limits_{j\in \mathcal{S}}\sum\limits_{u\in \Phi_{\mathrm{U},j}\backslash U_{0}}P^{\mathrm{UL}}_{j}l^{\mathrm{S}}_{j}(d_{u,j}^{\mathrm{UL}}),
\end{align}
where $U_{0}$ is the typical UE, $\Phi_{\mathrm{U},j}$ is the set of UE connected to the $j^{\mathrm{th}}$ tier,   and $d_{u,j}^{\mathrm{UL}}$ is the distance from UE $u$ in the $j^{\mathrm{th}}$ tier to the typical BS.
Given the assumption that  $\lambda_{\mathrm{U}}\gg \lambda_{k}$, there is a high probability that each BS serves at least one UE. Due to orthogonal time/frequency resource partitioning, each BS only serves one UE per resource block.
On each resource block, the locations of interfering UE can be approximated as those of BSs other than the typical BS \cite{ding2017uplink, novlan2013analytical}. Accordingly, the UL interference can be approximated as follows
\begin{align}
\label{eq:sub6_interference_UL_appro}
I_{s}^{\mathrm{S}, \mathrm{UL}}\approx \sum\limits_{j\in \mathcal{S}} \sum\limits_{i\in \Phi_{j}\backslash B_{s}^{\mathrm{UL}}}P^{\mathrm{UL}}_{j}l^{\mathrm{S}}_{j}(d_{i,j}),
\end{align}
where $B_{s}^{\mathrm{UL}}$ is the typical BS (UL serving BS) in the $s^{\mathrm{th}}$ sub-6 GHz tier.

The conditional coverage probability, when the typical UE connects to the $s^{\mathrm{th}}$ tier sub-6 GHz network, is derived using the SINR specified in (\ref{eq:sub6_SINR}).

\begin{theorem}
	\label{theorem_sub6_CCP}
	The conditional coverage probability for the DL and UL when the typical UE is connected to the $s^{\mathrm{th}}$ sub-6 GHz tier is given by 
	\begin{align}
	&\mathbb{P}(\mathrm{SINR}^{q}_{s}>\tau)  \nonumber\\
	&=\prod\limits_{j\in \mathcal{S}}\mathrm{exp}\left(-2\pi\lambda_j \int_{\varrho_{s,j}^{q}}^{\infty}\left(1- \frac{d_{i,j}^{\alpha_{j}}}{d_{i,j}^{\alpha_{j}} + Y_s^{q}(\tau)P_{j}^{q}}\right)r \mathrm{d}r\right) \times \nonumber\\ &\qquad\quad \mathrm{exp}\!\left( -Y_s^{q}(\tau)O^{\mathrm{S}}_s\right),
	\end{align}
	where $q\in\{\mathrm{DL, UL}\}$, $O^{\mathrm{S}}_s=\frac{\delta_{s}^2}{\beta_{0}}$, and $Y_s^{q}(\tau) = \frac{\tau x^{\alpha_{s}}}{P_{s}^{q}}$, and $\varrho_{s,j}^{q}=\left(\frac{C_j^{q}P_j^{q}}{C_s^{q}P_s^{q}} \right)^{\frac{1}{\alpha_{j}}}x^{\frac{\alpha_{s}}{\alpha_{j}}}$.	
\end{theorem}

\begin{IEEEproof}
	See Appendix \ref{App:sub6_CCP}.
\end{IEEEproof}

}

\subsubsection{mmWave SINR Coverage Probability}
%SINR of mmWave....
When the typical UE is connected to the $m^{\mathrm{th}}$ mmWave tier, with $D_{m}$ denoted as $x$, the DL and UL received SINR can be expressed as
\begin{align}
\label{eq:mmWave_SINR}
&\mathrm{SINR}^{q}_{m}(x) 
=\frac{P_{m}^{q}G_{m}^{\mathrm{max}}l^{\mathrm{M}}_{m}(x)}{I_{m}^{\mathrm{M},q} + \delta_{m}^2},
\end{align}
where $q\in\{\mathrm{DL, UL}\}$, $\delta_{m}^2$ is the AWGN power, $I_{m}^{\mathrm{M}, \mathrm{DL}}$ is the aggregated DL interference, which is given by
\begin{align}
\label{eq:mmWave_interference_DL}
I_{m}^{\mathrm{M}, \mathrm{DL}}=\sum\limits_{j\in \mathcal{M}} \sum\limits_{i\in \Phi_{j}^{\mathrm{L}}\backslash B_{m}^{\mathrm{DL}}}P^{\mathrm{DL}}_{j}G_{j}(N_j, \phi_{{D}_{i, j}})l^{\mathrm{M}}_{j}(d_{i,j}),
\end{align}
where  $B_{m}^{\mathrm{DL}}$ is the DL serving BS in the $m^{\mathrm{th}}$ mmWave tier, $\Phi^{\mathrm{L}}_{j}$ is the set of LOS BSs in the $j^{\mathrm{th}}$ tier, and $I_{m}^{\mathrm{M}, \mathrm{UL}}$ is the aggregated UL interference, which is given by
\begin{align}
\label{eq:mmWave_interference_UL}
I_{m}^{\mathrm{M}, \mathrm{UL}}=\sum\limits_{j\in \mathcal{M}}\sum\limits_{u\in \Phi_{\mathrm{U},j}^{\mathrm{L}}\backslash U_{0}}P^{\mathrm{UL}}_{j}G_{j}(N_j, \phi_{{D}_{u, j}^{\mathrm{UL}}})l^{\mathrm{M}}_{j}(d_{u,j}^{\mathrm{UL}}),
\end{align}
where $\Phi_{\mathrm{U},j}^{\mathrm{L}}$ is the set of LOS UE connected to the $j^{\mathrm{th}}$ tier, 
$\phi_{{D}_{u, j}^{\mathrm{UL}}}=\frac{1}{2}\left(\mathrm{cos}\phi_{U_{u,j}}-\mathrm{cos}\phi_{S_{u, j}^{\mathrm{UL}}}\right)$, 
$U_{u,j}$ denotes UE $u$ in the $j^{\mathrm{th}}$ tier, 
$\phi_{U_{u,j}}$ is the azimuth angle between  $U_{u,j}$ and the typical BS that serves the typical UE, and $\phi_{S_{u, j}^{\mathrm{UL}}}$ is the azimuth angle between $U_{u,j}$ and its serving BS.
Following the approximation in (\ref{eq:sub6_interference_UL_appro}), we have the following expression for the UL transmission in mmWave tiers
\begin{align}
\label{eq:mmWave_interference_UL_appro}
I_{m}^{\mathrm{M}, \mathrm{UL}}\approx \sum\limits_{j\in \mathcal{M}} \sum\limits_{i\in \Phi_{j}^{\mathrm{L}}\backslash B_{m}^{\mathrm{UL}}}P^{\mathrm{UL}}_{j}G_{j}(N_j, \phi_{{D}_{i, j}})l^{\mathrm{M}}_{j}(d_{i,j}),
\end{align}
where  {$B_{m}^{\mathrm{UL}}$ is the typical BS in the $m^{\mathrm{th}}$ mmWave tier}.

According to the predefined SINR in (\ref{eq:mmWave_SINR}), we derive the conditional coverage probability when the typical UE connects to an mmWave tier.

\begin{theorem}
	\label{theorem_mmWave_CCP}
	The conditional coverage probability for the DL and UL when the typical UE is connected to the $m^{\mathrm{th}}$ tier mmWave network is given by 
	\begin{align}
	&\mathbb{P}(\mathrm{SINR}^{q}_{m}>\tau)  \nonumber\\
	&\approx  \sum_{n=1}^{\gamma_m}(-1)^{n+1}\left( \substack{\gamma_m \\ \\ n}\right) \mathrm{exp}\left(-V_m^{q}(\tau)O^{\mathrm{M}}_m\right)  \times\nonumber\\ 
	&\prod\limits_{j\in\! \mathcal{M}}  \prod\limits_{w\in\! \{\mathrm{max},\mathrm{min}\}} \!\!\!\!\!\!\!\!\! \mathrm{exp}\!\left(-\!2\pi\lambda_jP_{\mathrm{G},j}^w\!\!\int_{\Omega_{m,j}^{q}}^{\infty} \!\!\!\!\!\! \left[ 1 \!-\! \Delta_{m,j}^{s,w}(r) \right] P_{\mathrm{LOS}}(r) r \mathrm{d}r \!\right) \!,
	\end{align}
 where $q\in\{\mathrm{DL, UL}\}$,  $P_{\mathrm{G},j}^{\mathrm{max}}=\phi_{\mathrm{3dB},j}/0.5=2\phi_{\mathrm{3dB},j}$ is the probability that the typical UE (BS) is located in the main-lobe direction, $P_{\mathrm{G},j}^{\mathrm{min}}=1-P_{\mathrm{G},j}^{\mathrm{max}}$ is the probability that the typical UE (BS) is located in the side-lobe direction,
 $O^{\mathrm{M}}_m=\delta_{m}^{2}\left(\frac{4\pi f_{\mathrm{M}}}{c}\right)^{2}$, $V_m^{q}(\tau) = \frac{n\eta_{m}\tau x^{\alpha_{m}}}{P_{m}^{q}N_{m}}$, $\eta_{m}=\gamma_m(\gamma_m !)^{-\frac{1}{\gamma_m}}$, $\Delta_{m,j}^{s,w}(r)= \left( 1 + \frac{V_m^{q}(\tau)P_{j}^{q}G_{j}^w r^{-\alpha_{j}}}{\gamma_j}  \right)^{-\gamma_j}$, and $\Omega_{m,j}^{q}=\left(\frac{C_j^{q}P_j^{q}N_j}{C_m^{q}P_m^{q}N_m} \right)^{\frac{1}{\alpha_{j}}}x^{\frac{\alpha_{m}}{\alpha_{j}}}$.	
\end{theorem}

\begin{IEEEproof}
	See Appendix \ref{App:mmWave_CCP}.
\end{IEEEproof}

\subsubsection{THz SINR Coverage Probability}
%SINR of THz...
With the typical UE connected to the $t^{\mathrm{th}}$ THz tier, denoting $D_{t}=x$, the DL and UL received SINR are given by
\begin{align}
\label{eq:Thz_SINR}
\mathrm{SINR}^{q}_{t}(x)
&=\frac{P_{t}^{q}G_{t}^{\mathrm{max}}l^{\mathrm{T}}_{t}(x)}{I_{t}^{\mathrm{T},q} + \mathrm{Noise}_{t}^{q}(x)},
\end{align}
where $q\in\{\mathrm{DL, UL}\}$, $I_{t}^{\mathrm{T},q}$ is the aggregated interference, and $\mathrm{Noise}_{t}^{q}$ is the cumulative molecular absorption and thermal noises. {The interference signals only come from the THz tiers.} More specifically, the DL aggregated interference is given by
\begin{align}
\label{eq:interference_t}
&I_{t}^{\mathrm{T},\mathrm{DL}} \nonumber\\
&= \sum\limits_{j\in  \mathcal{T}} \sum\limits_{i\in \Phi_{j}^{\mathrm{L}}\backslash B_{t}^{\mathrm{DL}}}P^{\mathrm{DL}}_{j}G_{j}(N_j, \phi_{{D}_{i, j}})l^{\mathrm{M}}_{j}(d_{i,j}) \nonumber\\
&= \!\!\sum\limits_{j\in  \mathcal{T}} \sum\limits_{i\in \Phi_{j}^{\mathrm{L}}\backslash B_{t}^{\mathrm{DL}}}\!\! P^{\mathrm{DL}}_{j}G_{j}(N_j, \phi_{{D}_{i, j}})\!\!\left(\frac{c}{4\pi f_{\mathrm{T}}}\right)^{\!2\!}\!\!d_{i,j}^{-\alpha_{t}}e^{-\!K_ad_{i,j}}, 
\end{align}
where $B_{t}^{\mathrm{DL}}$ is the DL serving BS in the $t^{\mathrm{th}}$ THz tier. The DL total noise power is given by
\begin{align}
\label{eq:noise_t}
&\mathrm{Noise}_{t}^{\mathrm{DL}}(x) \nonumber\\
&= \!\!\sum\limits_{j\in \mathcal{T}} \sum\limits_{i\in \Phi_{j}^{\mathrm{L}}\backslash B_{t}^{\mathrm{DL}}}\!\!\!P^{\mathrm{DL}}_{j}G_{j}\!\!\left(\!N_j, \phi_{{D}_{i, j}}\!\right)\!\left(\!\frac{c}{4\pi f_{\mathrm{T}}}\!\right)^{\!2\!}\!d_{i,j}^{-\alpha_{t}}\!\left(1\!-\!e^{-\!K_ad_{i,j}}\!\right)  \nonumber\\
&\ +P_{t}^{\mathrm{DL}}G_{t}^{\mathrm{max}}\left(\frac{c}{4\pi f_{\mathrm{T}}}\right)^{2}x^{-\alpha_{t}}\left(1-e^{-K_ax}\right)  + \delta_{t}^2, 
\end{align}
where the first two terms represent the molecular absorption noise from the interference signals and the desired signal, respectively, and $\delta_{t}^2$ is the Johnson-Nyquist noise. 
{ The power spectral density (PSD) of $\delta_{t}^2$ remains constant up to 0.1 THz at $P_{\mathrm{JN}}=k_{\mathrm{B}}T_{0}=-174$ dBm/Hz, where $k_{\mathrm{B}}$ is the Boltzmann constant and $T_{0}$ is the temperature in Kelvin. When $f_{\mathrm{T}}>0.1$ THz, the Johnson-Nyquist noise becomes frequency-dependent and the PSD can be computed as $P_{\mathrm{JN}}=\frac{f_{\mathrm{T}}p}{\mathrm{exp}\left(\frac{f_{\mathrm{T}}p}{k_{\mathrm{B}}T_{0}}\right)-1}$ dBm/Hz, where $p$ is Planck's constant \cite{Petrov2015Interference}. Accordingly, we have $\delta_{t}^2=P_{\mathrm{JN}}B_{\mathrm{W},t}$,
where $B_{\mathrm{W},t}$ is the bandwidth of the $t^{\mathrm{th}}$ THz tier.}
Combining (\ref{eq:interference_t}) and (\ref{eq:noise_t}), we have
\begin{align}
&I_{t}^{\mathrm{T},\mathrm{DL}} +  \mathrm{Noise}_{t}^{\mathrm{DL}}(x)  \nonumber\\
&= \sum\limits_{j\in  \mathcal{T}} \sum\limits_{i\in \Phi_{j}^{\mathrm{L}}\backslash B_{t}^{\mathrm{DL}}}P^{\mathrm{DL}}_{j}G_{j}\left(N_j, \phi_{{D}_{i, j}}\right)\left(\frac{c}{4\pi f_{\mathrm{T}}}\right)^{2}d_{i,j}^{-\alpha_{t}} +   \nonumber\\
&\quad\ P_{t}^{\mathrm{DL}}G_{t}^{\mathrm{max}}\left(\frac{c}{4\pi f_{\mathrm{T}}}\right)^{2}x^{-\alpha_{t}}\left(1-e^{-K_ax}\right)  +  \delta_{t}^2.
\end{align}

According to the approximation provided in (\ref{eq:sub6_interference_UL_appro}), the expression for the UL transmission can be derived as
\begin{align}
&I_{t}^{\mathrm{T},\mathrm{UL}} +  \mathrm{Noise}_{t}^{\mathrm{UL}}(x)   \nonumber\\
&\approx \sum\limits_{j\in  \mathcal{T}} \sum\limits_{i\in \Phi_{j}^{\mathrm{L}}\backslash B_{t}^{\mathrm{UL}}}P_{j}^{\mathrm{UL}}G_{j}\left(N_j, \phi_{{D}_{i, j}}\right)\left(\frac{c}{4\pi f_{\mathrm{T}}}\right)^{2}d_{i,j}^{-\alpha_{t}} \,\, +   \nonumber\\
&\quad P_{t}^{\mathrm{UL}}G_{t}^{\mathrm{max}}\left(\frac{c}{4\pi f_{\mathrm{T}}}\right)^{2}x^{-\alpha_{t}}\left(1-e^{-K_ax}\right)  +  \delta_{t}^2,
\end{align}
 {where $B_{t}^{\mathrm{UL}}$ is the typical BS in the $t^{\mathrm{th}}$ THz tier.}

When the typical UE is connecting to the $t^{\mathrm{th}}$ tier THz network, the conditional coverage probability can be derived based on the SINR in (\ref{eq:Thz_SINR}).
\begin{theorem}
	\label{theorem_Thz_CCP}
	The conditional coverage probability for the DL and UL when the typical UE is connected to the $t^{\mathrm{th}}$ tier THz network is given by
	\begin{align}
	&\mathbb{P}(\mathrm{SINR}^{q}_{t}(x)>\tau)  \nonumber\\
	&\approx \!\!\sum_{n=1}^{\gamma_{\mathrm{T}}}\!(-1)^{n+1}\!\!\left( \substack{\gamma_{\mathrm{T}} \\ \\ n}\right) \!\mathrm{exp}\!\Bigg(\!\! - \! n\eta_{\mathrm{T}}\tau \bigg(\!\frac{\delta_{t}^2 e^{K_ax}}{J_t^{q}(x)} + e^{K_ax}\!-\!\! 1 \!\!\bigg)\!\Bigg)\!  \nonumber\\ 
	&\quad\times \prod\limits_{j\in  \mathcal{T}} \prod\limits_{w\in \{\mathrm{max},\mathrm{min}\}} \mathrm{exp}\Bigg(	 -2\pi\lambda_jP_{\mathrm{G},j}^w \int_{\Theta_{t,j}^{q}}^{\infty} P_{\mathrm{LOS}}(r)r  \nonumber\\
	&\quad\times \Bigg( 1- \mathrm{exp}\bigg( - n\eta_{\mathrm{T}}\tau x^{\alpha_{t}}e^{K_ax}  \frac{P_{j}^{q}G_{j}^{w}}{P_{t}^{q}N_{t}}r^{-\alpha_{t}}\bigg)\Bigg) \mathrm{d}r \Bigg),
	\end{align}
	where $\gamma_{\mathrm{T}}$ is the shape parameter of the induced Nakagami-$m$ distribution $h_{\mathrm{T}}\sim\Gamma(\gamma_\mathrm{T},\frac{1}{\gamma_\mathrm{T}})$ (when $\gamma_{\mathrm{T}}\rightarrow \infty$, $h_{\mathrm{T}}\rightarrow 1$), $\eta_{\mathrm{T}}=\gamma_{\mathrm{T}}(\gamma_{\mathrm{T}} !)^{-\frac{1}{\gamma_{\mathrm{T}}}}$, $J_t^{q}(x)=P_{t}^{q}G_{t}^{\mathrm{max}}\left(\frac{c}{4\pi f_{\mathrm{T}}}\right)^{2}x^{-\alpha_{t}}$, and $\Theta_{t,j}^{q}=\frac{\alpha_{j}}{K_a}W_{\mathrm{L}}(\Lambda_{t,j}^{\mathrm{T},q}e^{\frac{K_a}{\alpha_{j}}x}x^{\frac{\alpha_{t}}{\alpha_{j}}})$, where $\Lambda_{t,j}^{\mathrm{T},q}$ is defined in Lemma 4.	
\end{theorem}

\begin{IEEEproof}
	See Appendix \ref{App:Thz_CCP}.
\end{IEEEproof}

\subsection{Rate Coverage Probability}
Due to the abundant bandwidths, mmWave/THz frequency bands can provide ultra-high data rate. In this respect, rate coverage probability is a crucial performance metric that assesses the capacity of a wireless network to ensure reliable communication at a specific data rate over a certain geographic region. The DL (UL) rate coverage probability is defined as the probability that the DL (UL) rate is higher than a given threshold as follows
\begin{align}
\label{eq:RC}
R_{\mathrm{cov}}^{q}(\rho) &= \sum_{k=1}^{K}R_{\mathrm{cov},k}^{q}(\rho)\mathcal{A}_{k}^{q},
\end{align}
where $q\in\{\mathrm{DL, UL}\}$, $\rho$ is the rate threshold, and $R_{\mathrm{cov},k}^{q}(\rho)$ is the conditional rate coverage probability when the typical UE is associated with the $k^{\mathrm{th}}$ tier. The conditional rate coverage probability can be further derived as
\begin{align}
\label{eq:RC_k}
R_{\mathrm{cov},k}^{q}(\rho) &= \mathbb{E}_{x}\left[\mathbb{P}\left[\frac{B_{\mathrm{W},k}}{Z_{k}^{q}}\mathrm{log}_{2}(1+\mathrm{SINR}_{k}^{q}(x))>\rho\right]\right] \nonumber\\
&= \mathbb{E}_{x}\left[\mathbb{P}\left[\mathrm{SINR}_{k}^{q}(x)>2^{\frac{\rho Z_{k}^{q}}{B_{\mathrm{W},k}}}-1\right]\right] 
\nonumber\\
&={P_{\mathrm{cov},k}^{q}\left(2^{\frac{\rho Z_{k}^{q}}{B_{\mathrm{W},k}}}\!-\!1\right),}
\end{align}
where $B_{\mathrm{W},k}$ is the bandwidth of the $k^{\mathrm{th}}$ tier and $Z_{k}^{q}$ is the average number of UE served by a BS in the $k^{\mathrm{th}}$ tier, which is given by \cite{singh2013offloading}
\begin{align}
Z_{k}^{q} = 1 + \frac{1.28\lambda_{\mathrm{U}}\mathcal{A}_{k}^{q}}{\lambda_{k}}.
\end{align}

According to (\ref{eq:RC}) and (\ref{eq:RC_k}), the rate coverage probability for the DL and UL can be derived as 
{
\begin{align}
&R^{q}_{\mathrm{cov}}(\rho)  \nonumber\\
&=\sum_{s\in \mathcal{S}} \! R^{q}_{\mathrm{cov},s}(\rho)\mathcal{A}^{q}_{s}  + \!\!\! \sum_{m\in \mathcal{M}}R^{q}_{\mathrm{cov},m}(\rho)\mathcal{A}^{q}_{m}  + \! \sum_{t\in \mathcal{T}}R^{q}_{\mathrm{cov},t}(\rho)\mathcal{A}^{q}_{t} \nonumber\\
&=\sum_{s\in \mathcal{S}}P_{\mathrm{cov},s}^{q} \! \left(2^{\frac{\rho Z_{s}^{q}}{B_{\mathrm{W},s}}}\!-\!1\right)\mathcal{A}^{q}_{s}  \!+\!\!\!  \sum_{m\in \mathcal{M}}\!\!P_{\mathrm{cov},m}^{q} \! \left(2^{\frac{\rho Z_{m}^{q}}{B_{\mathrm{W},m}}}\!-\!1\right)\mathcal{A}^{q}_{m}  \nonumber\\ 
&\quad\ +\sum_{t\in \mathcal{T}}P_{\mathrm{cov},t}^{q} \! \left(2^{\frac{\rho Z_{t}^{q}}{B_{\mathrm{W},t}}}\!-\!1\right)\mathcal{A}^{q}_{t}.
\end{align}
}

\section{Numerical Results}
\label{sec:numerical}
In this section, we validate the derived analytical expressions with Monte Carlo simulations. Each simulation consists of $5\times 10^{4}$ independent random realisations according to the system model described in Section \ref{sec:system_model}. We consider a three-tier hybrid {sub-6GHz-mmWave-THz} network by default.
The default numerical simulation parameter values are listed in Table \ref{tab2} unless otherwise stated \cite{chen2022deployment,hossan2021mobility,Chen2021Optimal,Elshaer2016Downlink,Zhang2020Energy}, {where subscript 1 denotes the sub-6 GHz tier,  subscript 2 denotes the mmWave tier and subscript 3 denotes the THz tier.} The impacts of different system parameters on the association probability, SINR coverage probability and rate coverage probability are investigated.

\begin{table}[t]
	\caption{Value of Parameters}
	\begin{center}
 \scalebox{1}{
		\begin{tabular}{|c|c|}
			\hline   
			\textbf{Parameters}&\textbf{Default value}\\
			\hline  \hline
			$\lambda_{\mathrm{b}}$, $\lambda_{\mathrm{U}}$ &  $10^{-3}$ $\mathrm{m}^{-2}$, $2\times 10^{-3}$ $\mathrm{m}^{-2}$  \\
			\hline 
			$L$, $W$ & 15 m, 15 m \\
			\hline
			\tabincell{c}{{$\lambda_{1}$}, $\lambda_{2}$, $\lambda_{3}$} & \tabincell{c}{$2\times10^{-6}$ $\mathrm{m}^{-2}$, $5\times 10^{-5}$ $\mathrm{m}^{-2}$, \\ $5\times10^{-4}$ $\mathrm{m}^{-2}$ } \\
			\hline
			$N_{2}$, $N_{3}$ &  64, 100 \\
			\hline
			$\alpha_{1}$, $\alpha_{2}$, $\alpha_{3}$ & 4, 2, 2 \\
			\hline 
			$\gamma_{2}$, $\gamma_{\mathrm{T}}$ & 3, 10  \\
			\hline
			{$P_{1}^{\mathrm{DL}}$}, $P_{2}^{\mathrm{DL}}$, $P_{3}^{\mathrm{DL}}$ & 46 dBm, 33 dBm, 23 dBm \\
			\hline
            {$P_{1}^{\mathrm{UL}}$}, $P_{2}^{\mathrm{UL}}$, $P_{3}^{\mathrm{UL}}$ & 23  dBm, 23 dBm, 23 dBm  \\
			\hline
			$f_{\mathrm{M}}$, $f_{\mathrm{T}}$ & 28 GHz, 340 GHz \\
			\hline
			$\beta_{0}$ & -38.5 dB  \\
			\hline
			\tabincell{c}{$C_{1}^{q}$, $C_{2}^{q}$, $C_{3}^{q}$, \\ $q\in\{\mathrm{DL, UL}\}$} & 1, 1, 1 \\
			\hline
            $K_a$ & 0.01 \\
			\hline
			{$\delta_{1}^2$}  & \tabincell{c}{ $-174 \, \mathrm{dBm/Hz} + 10\mathrm{log}_{10}(B_{\mathrm{W},1})$ \\ $+ 10 \, \mathrm{dB}$}  \\
   \hline
			$\delta_{2}^2$  & \tabincell{c}{ $-174 \, \mathrm{dBm/Hz} + 10\mathrm{log}_{10}(B_{\mathrm{W},2})$ \\ $+ 10 \, \mathrm{dB}$}  \\
			\hline
			{$B_{\mathrm{W},1}$}, $B_{\mathrm{W},2}$, $B_{\mathrm{W},3}$  &  10 MHz, 1 GHz, 10 GHz  \\
            \hline
            {$p$}  &  $6.62607015\times10^{-34} \mathrm{J\cdot S}$   \\
            \hline
            $\tau$, $\rho$  &  10 dB, $10^{9}$ bit/s  \\ 
			\hline
		\end{tabular}}
		\label{tab2}
	\end{center}
\end{table}

\subsection{Association Probability}

\begin{figure}[t]
\centerline{\includegraphics[width=3.4in]{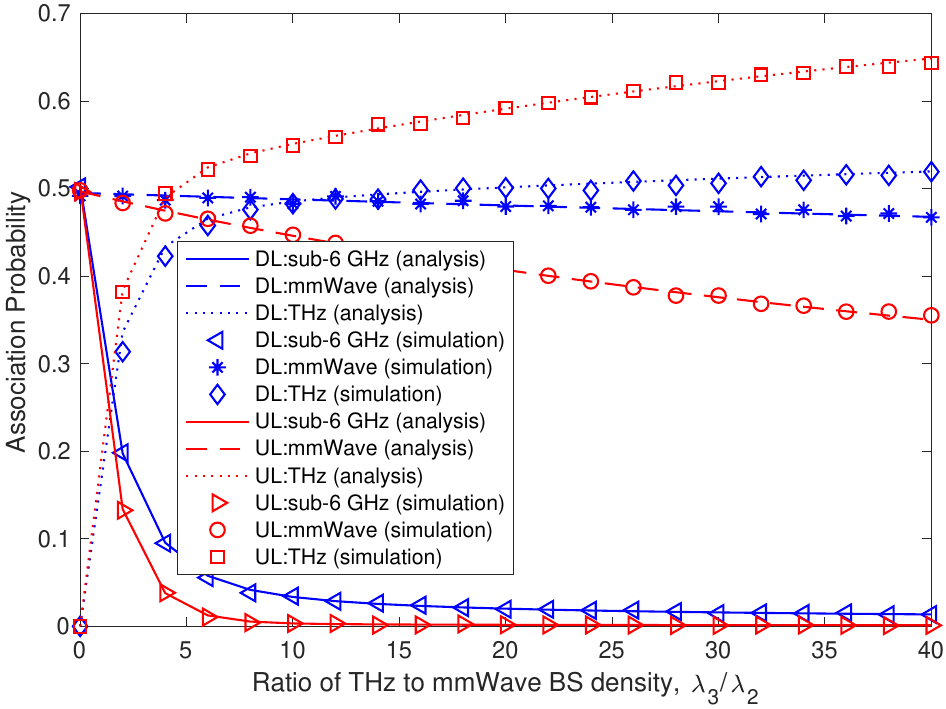}}
\caption{{The analytical and simulation results of the association probability versus the ratio of THz  to mmWave BS density.}}
\label{fig:AP_BSden_AS}
\end{figure}

Fig. \ref{fig:AP_BSden_AS} presents the DL and UL association probabilities versus the ratio of THz  to mmWave BS density. It is clear that the analytical results  align well with the simulation results, validating the correctness of our derived theoretical expressions. The THz tier benefits from the increase of THz BSs for both DL and UL  transmissions. On the other hand, a notable difference between DL and UL association probabilities is observed, indicating that the typical UE is inclined to connect to different BSs during DL and UL associations. This disparity increases with the density ratio of THz BSs to mmWave BSs. 

\begin{figure}[t]
\centerline{\includegraphics[width=3.4in]{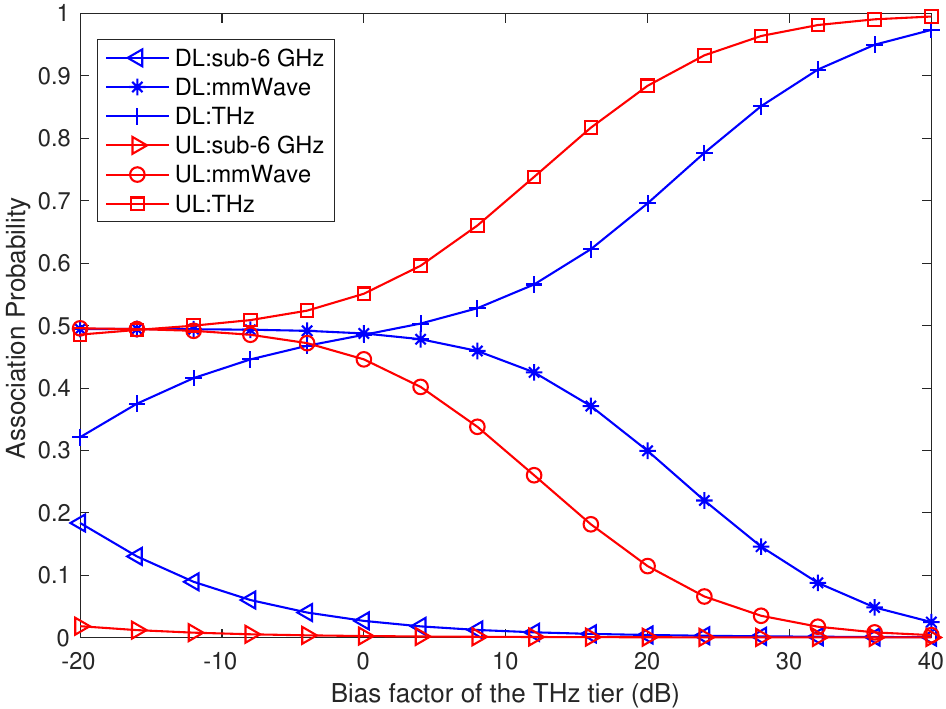}}
\caption{{The analytical results of the association probability versus the bias factor of the THz tier.}}
\label{fig:AP_Bias_A}
\end{figure}

In Fig. \ref{fig:AP_Bias_A}, the DL and UL association probabilities are plotted versus the bias factor of the THz tier. It is observed that the association probability of the THz tier monotonically increases with the bias factor. This result is quite intuitive because more UE are encouraged to connect with the THz BSs when the bias factor of the THz tier increases. Another observation is that the probability of connecting to the {sub-6 GHz}/mmWave tier in the DL is always higher than that in the UL, as the DL transmit power of {sub-6 GHz}/mmWave BSs is higher than that of THz BSs.

\subsection{SINR and Rate Coverage Probabilities}

\begin{figure}[t]
\centerline{\includegraphics[width=3.4in]{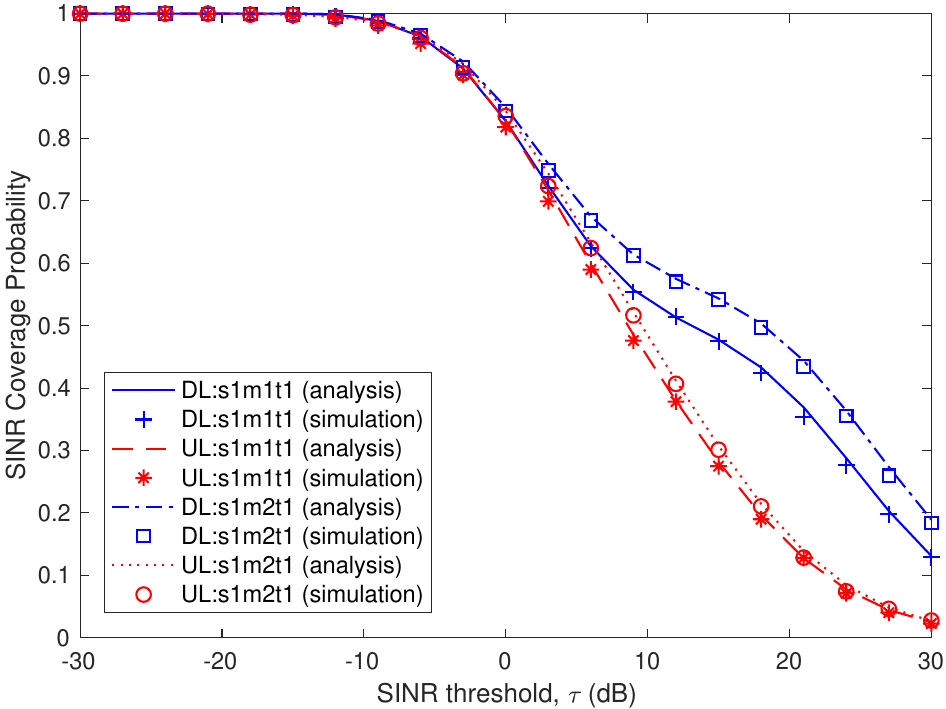}}
\caption{{The analytical and simulation results of the SINR coverage probability versus the SINR threshold. We present results of a 3-tier hybrid network (s1m1t1) consisting of 1 tier of sub-6 GHz network, 1 tier of mmWave network and 1 tier of THz network, and a 4-tier hybrid network (s1m2t1) consisting of 1 tier of sub-6 GHz network, 2 tiers of mmWave networks and 1 tier of THz network.}}
\label{fig:CP_Valid_AS}
\end{figure}

Fig. \ref{fig:CP_Valid_AS} shows the DL and UL SINR coverage probabilities versus the SINR threshold, where we present results for 3-tier and 4-tier hybrid sub-6GHz-mmWave-THz networks, respectively. For the 4-tier hybrid network, the $2^{\mathrm{nd}}$ mmWave tier  has the following simulation parameters: $\lambda_2 = 10^{-5} \ \mathrm{m}^{-2}$, $N_2= 32$, $P_2^\mathrm{DL} = 43$ dBm, and $P_2^\mathrm{UL} = 23$ dBm. The $1^{\mathrm{st}}$ sub-6 GHz tier, the $3^{\mathrm{th}}$ mmWave tier and the $4^{\mathrm{th}}$ THz tier have the same simulation parameters as the default values listed in Table \ref{tab2}.
The figure demonstrates that the analytical expressions provide accurate results that closely match the simulation curves. Thanks to the extra deployment of an mmWave tier, the 4-tier hybrid network has a higher SINR coverage probability than the 3-tier network. 
{The SINR coverage probability gain is more significant in the DL than in the UL due to the larger DL transmit power.}

\begin{figure}[t]
\centerline{\includegraphics[width=3.4in]{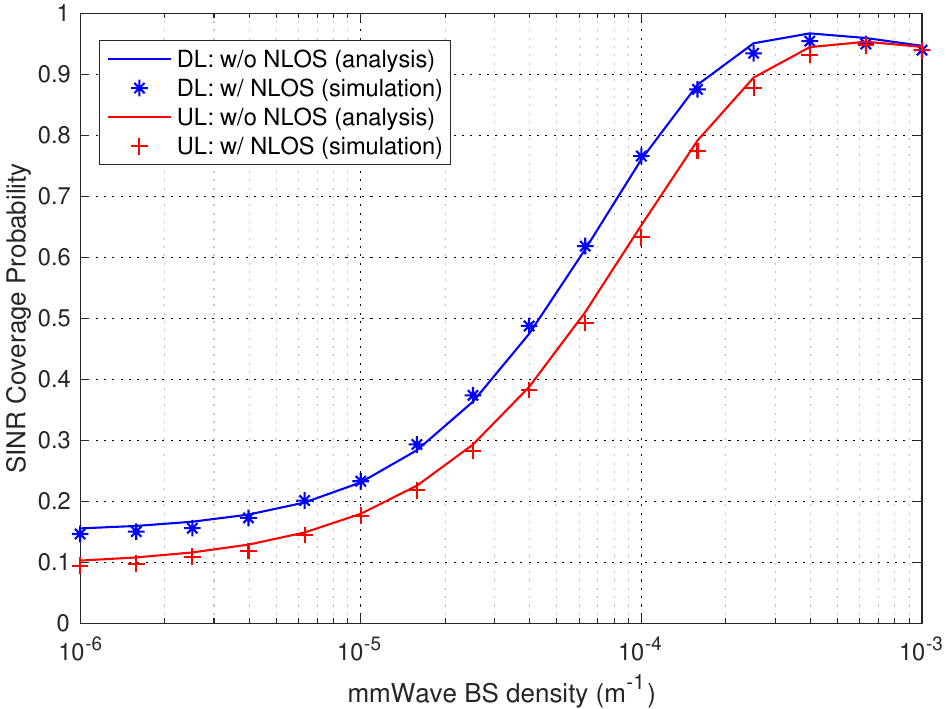}}
\caption{{The analytical and simulation results of the impact of NLOS transmission links on
the SINR coverage probability.}}
\label{fig:CP_NLOS_AS}
\end{figure}

 In Fig. \ref{fig:CP_NLOS_AS}, we justify the assumption of ignoring  NLOS transmission links stated in Section \ref{sec:channel_model}. We take into account the NLOS signals and interference of the mmWave tier.
The simulation parameters for NLOS propagation are set as follows: the path loss exponent is $\alpha_{2}^{\text{NLOS}}=4$, the intercept  is -72 dB, and the shape parameter of the small-scale fading power gain is $\gamma_2^{\text{NLOS}}=2$ \cite{bai2014coverage}. 
We can see that the curve incorporating NLOS transmission links aligns well with that ignoring NLOS components. This indicates that the impact of NLOS transmission links on the SINR coverage probability is negligible in the considered sub-6GHz-mmWave-THz network.

\begin{figure}[t]
\centerline{\includegraphics[width=3.4in]{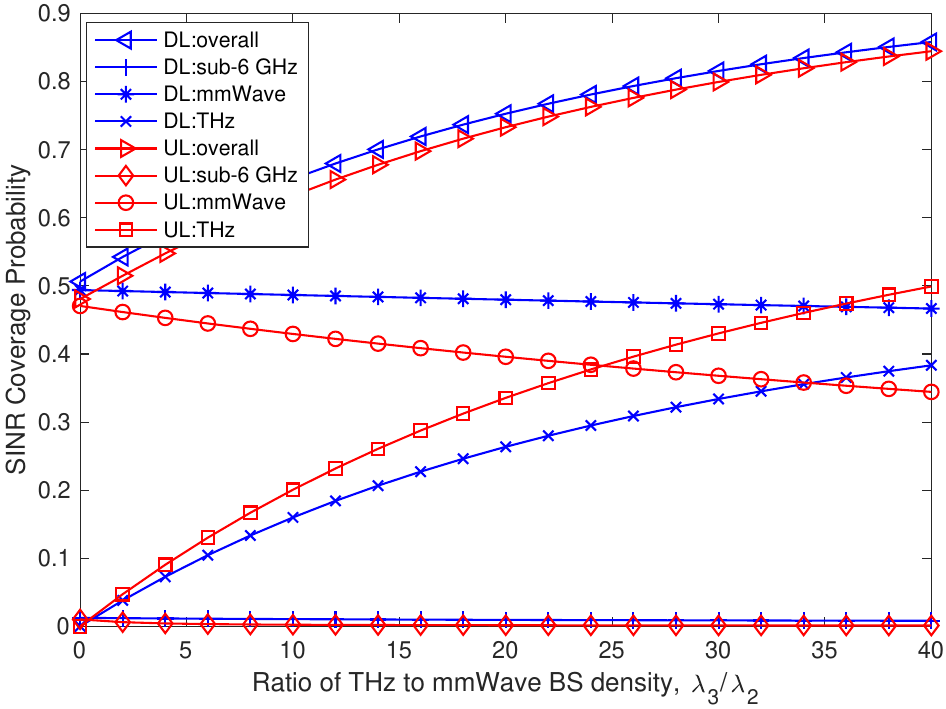}}
\caption{{The analytical results of the SINR coverage probability versus the ratio of THz to mmWave BS density with $\tau=5$ dB.}}
\label{fig:CP_BSden_A}
\end{figure}

In Fig. \ref{fig:CP_BSden_A}, the DL and UL SINR coverage probabilities are presented versus the ratio of THz  to mmWave BS density. We can see that the overall SINR coverage probability increases with the THz BS density due to the increased probability of connecting to an LOS BS and the reduced THz propagation loss. Due to the higher transmission power, the contribution of mmWave tier to the SINR coverage probability in the DL transmission is higher than that in the UL transmission.

\begin{figure}[t]
\centerline{\includegraphics[width=3.4in]{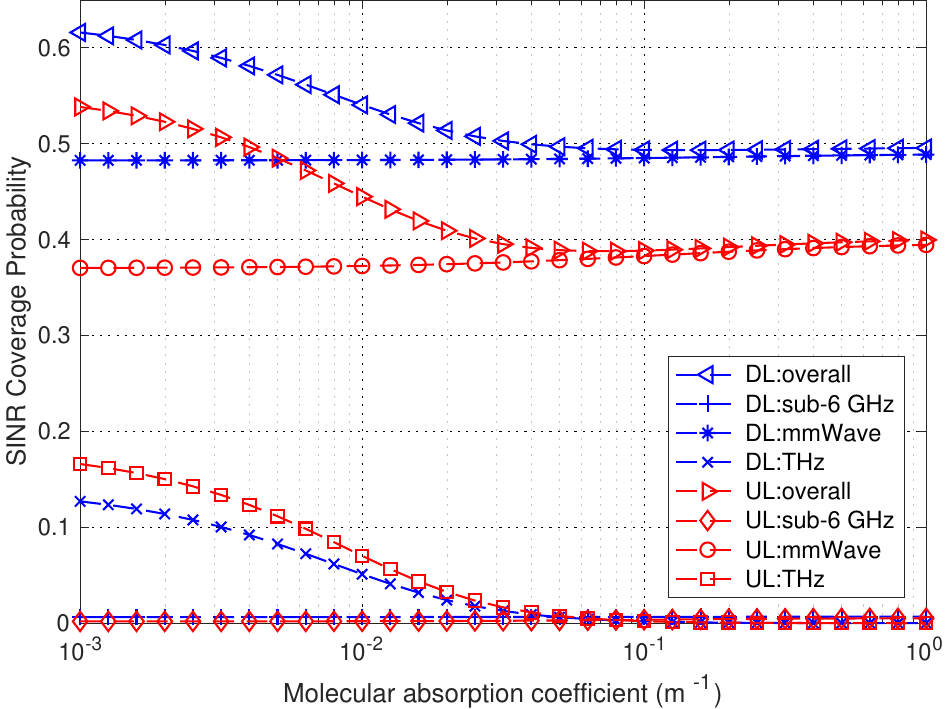}}
\caption{{The analytical results of the SINR coverage probability versus the molecular absorption coefficient.}}
\label{fig:CP_Ka_A}
\end{figure}

In Fig. \ref{fig:CP_Ka_A}, the DL and UL SINR coverage probabilities are depicted versus the molecular absorption coefficient of the THz tier. Note that our theoretical expressions are valid for any molecular absorption coefficient value $K_a$, which depends on the transmission environment and signal frequency. As can be observed, a higher molecular absorption coefficient generally leads to the decay of the SINR coverage probability. When $K_a>0.1  \ \mathrm{m}^{-1}$, the SINR coverage probability of the THz tier approaches zero as  all UE preferentially connect to the mmWave tier. This results in a slight increase in the overall SINR coverage probability due to more favourable propagation conditions in the mmWave layer.

\begin{figure}[t]
\centerline{\includegraphics[width=3.4in]{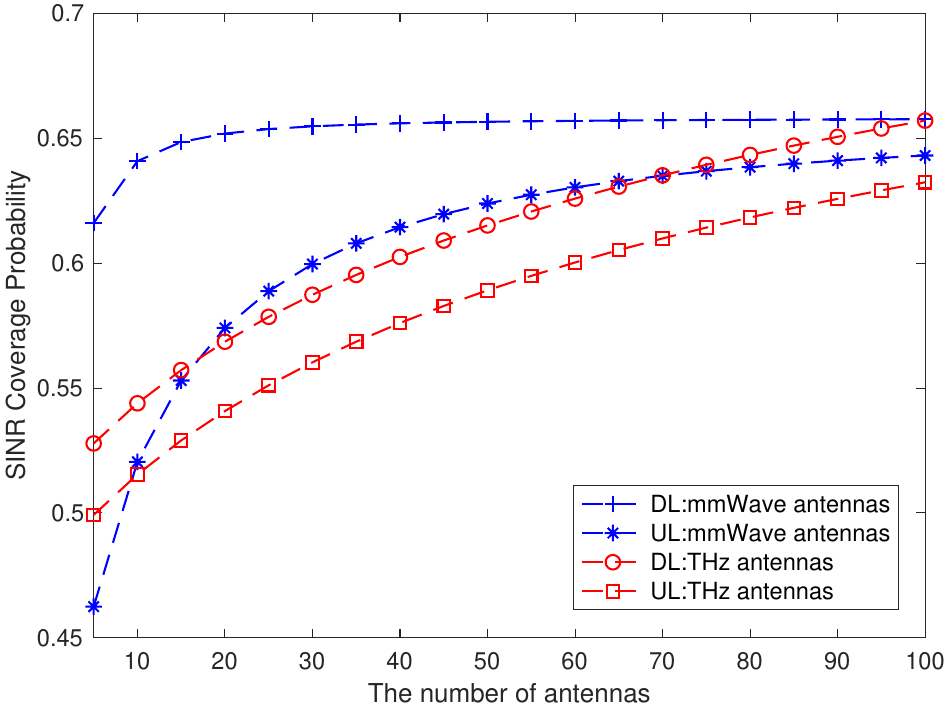}}
\caption{{The analytical results of the SINR coverage probability versus the number of antennas per BS at the mmWave/THz tier with $\tau=5$ dB.}}
\label{fig:CP_An}
\end{figure}

Fig. \ref{fig:CP_An} presents the impact of the number of antennas on the SINR coverage probability. It is observed that both DL and UL SINR coverage probabilities increase with the number of antennas per BS at the mmWave/THz tier. We see that the SINR coverage probability gain of increasing mmWave antennas in the UL is more significant than in the DL. This is because the UL transmit power in the mmWave tier is smaller than the DL transmit power. Hence, the increase of mmWave antennas in the UL significantly increases the probability of  connecting to the mmWave tier.

\begin{figure}[t]
\centerline{\includegraphics[width=3.4in]{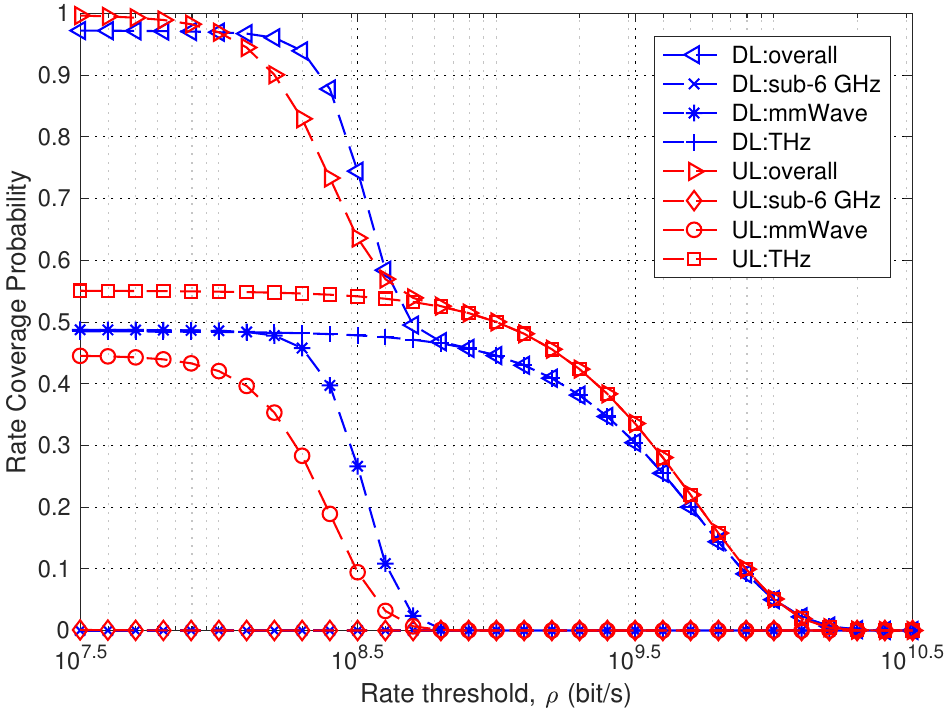}}
\caption{{The analytical results of the rate coverage probability versus the rate threshold.}}
\label{fig:RP_Thre}
\end{figure}

Fig. \ref{fig:RP_Thre} presents the DL and UL rate coverage probabilities against the rate threshold.  As demonstrated in the figure, due to the limited bandwidth, the sub-6 GHz tier and mmWave tier fail to contribute any rate coverage probability when the rate threshold is above $10^{9}$ bit/s, indicating that only the THz tier is capable of providing a rate higher than $10^{9}$ bit/s.

\subsection{Effect of DL and UL Decoupled Cell Association}
In this subsection, we show the effect of the bias factor on the SINR and rate coverage, and illustrate the necessity of using DL and UL decoupled cell-association strategy.

\begin{figure}[!t]
 \centering
 \subfloat[]{\includegraphics[width=3.4in]{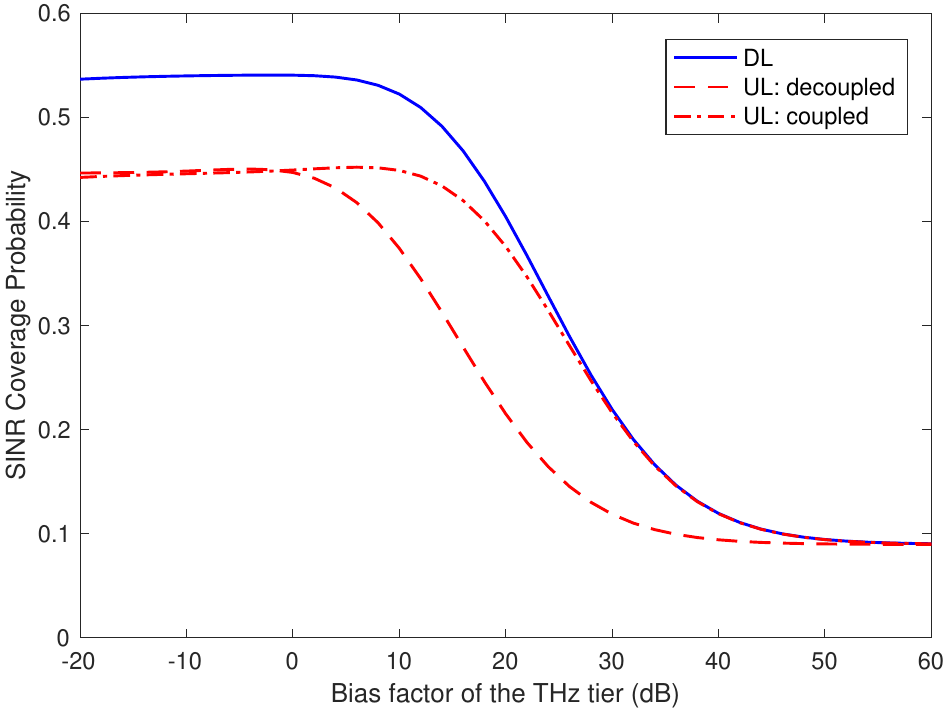}} \ \ \ \ \ 
 \subfloat[]{\includegraphics[width=3.4in]{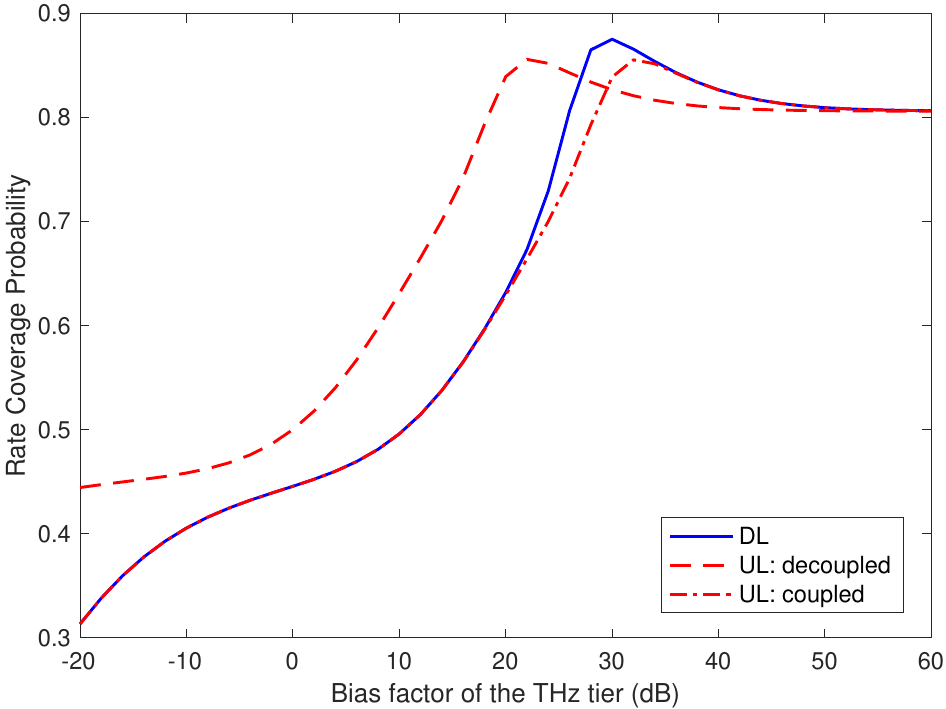}}
 \caption{{The analytical results of the SINR and rate coverage probabilities versus the bias factor of the THz tier. (a) SINR coverage probability. (b) Rate coverage probability.}}
\label{fig:CP_Bias_A}
\end{figure}

In Fig. \ref{fig:CP_Bias_A}, we show the SINR and rate coverage probabilities versus the bias factor of the THz tier, where $P_{k}^{\mathrm{UL}}$ denotes the UL transmit power in the $k^{\mathrm{th}}$ tier and $P_{k}^{\mathrm{UL}}=P_{k^{'}}^{\mathrm{UL}}, \forall k^{'}\in \mathcal{K}$ and $k^{'}\ne k$. In the DL and UL coupled cell-association strategy, the typical UE connects to the BS providing the strongest DL average biased received power for both DL and UL communications. 
From Fig. \ref{fig:CP_Bias_A}(a), 
it is observed that the SINR coverage probability first slightly increases and then rapidly decreases with the bias factor of the THz tier for all the cases considered. The initial increase comes from the increase of the contribution of the THz tier when the probability of connecting to the THz tier increases.
On the other hand, the THz tier suffers from a high penetration loss, as a result of which, a further increased bias factor lead to the decay of the SINR coverage. 
From Fig. \ref{fig:CP_Bias_A}(b), we see that the rate coverage probability first rapidly increases and then slowly decreases with the bias factor of the THz tier. 
UE  can benefit from a larger bandwidth of the THz band when the bias factor increases. Nevertheless, when the bias factor exceeds a critical threshold, the rate coverage suffers from a low SINR and excessive UE per THz BS. 
Generally, a larger bias factor of the THz tier brings a higher rate coverage probability, but at the cost of a lower SINR coverage probability. Moreover, we see that the coupled cell association strategy cannot achieve the maximum rate coverage for  DL and UL
transmissions simultaneously. In contrast, this can be achieved
by the DL and UL decoupled cell-association strategy due to the separately designed bias factors.

\begin{figure*}[!t]
 \centering
 \subfloat[]{\includegraphics[width=3.4in]{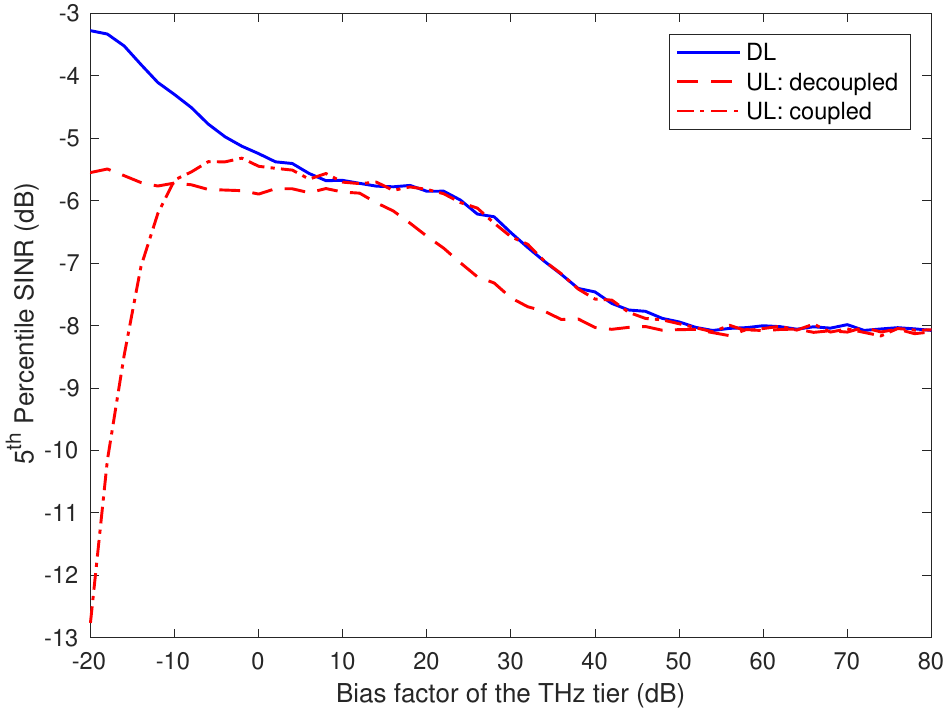}} \ \ \ \ \ 
 \subfloat[]{\includegraphics[width=3.4in]{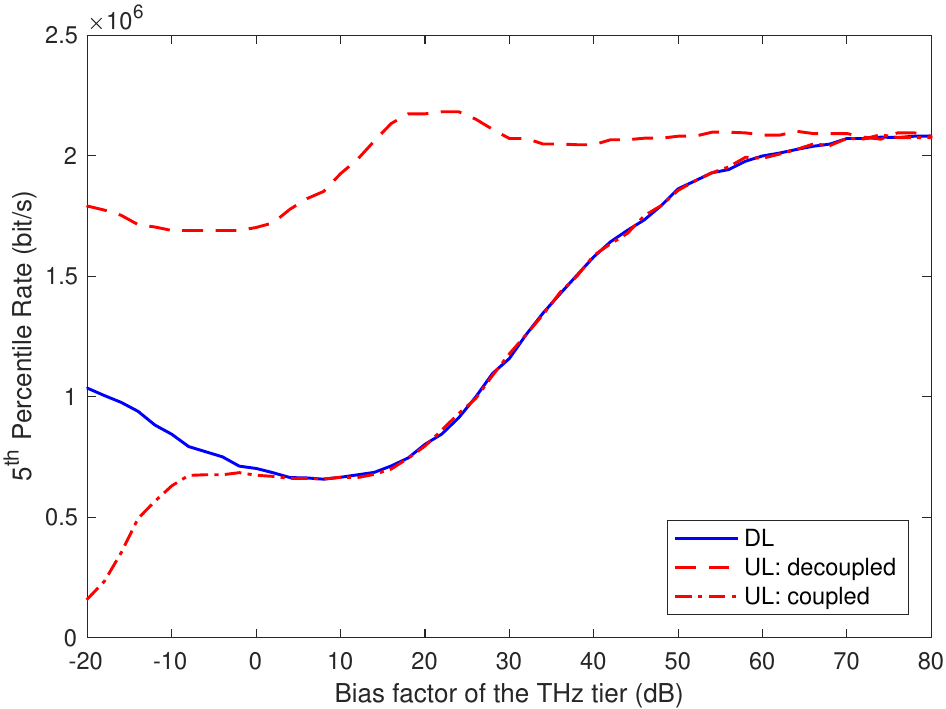}}
 \caption{{The simulation results of the $5^{\mathrm{th}}$ percentile SINR and rate  versus the bias factor of the THz tier. (a) $5^{\mathrm{th}}$ percentile SINR. (b) $5^{\mathrm{th}}$ percentile rate.}}
\label{fig:RP_Bias_A}
\end{figure*}

{In Fig. \ref{fig:RP_Bias_A}, we plot the $5^{\mathrm{th}}$ percentile SINR and rate versus the bias factor of the THz tier. 
These results can reflect the network performance experienced by cell-edge UE.
Similar with Fig. \ref{fig:CP_Bias_A}, we can observe that a large bias factor generally leads to a high rate but a low SINR.
From Fig. \ref{fig:RP_Bias_A}(b), we see that for small and medium bias factors, the decoupled UL case can achieve a significantly higher $5^{\mathrm{th}}$ percentile rate than the DL and coupled UL cases, as more UE are connected to the THz tier in the decoupled UL case.
In the decoupled UL case, a bias factor of $10$ dB can lead to good $5^{\mathrm{th}}$ percentile SINR and rate simultaneously.
However, this can not be achieved in the coupled UL case,
 indicating the necessity of using DL and UL decoupled cell association strategy.}

From Fig. \ref{fig:CP_Bias_A} and Fig. \ref{fig:RP_Bias_A}, we can summarise two important system design insights. First, there exists a trade-off between SINR and rate when choosing the cell-association bias factor.
The optimal bias factor that maximises the rate coverage generally results in a low SINR coverage, which leads to the need for robust modulation and coding techniques. Second, the DL and UL decoupled cell-association strategy allows a more flexible configuration of the bias factor than its coupled counterpart. This flexibility enables good average network performance and cell-edge network performance in both DL and UL communications.

\section{Conclusions}

In this paper, we have proposed a novel and tractable stochastic geometry-based framework for the performance evaluation of a general multi-tier hybrid sub-6GHz-mmWave-THz network. We have investigated the DL and UL decoupled cell-association strategy that allows the separate cell access during the DL and UL transmissions. Under the DL and UL decoupled cell-association strategy, we have derived novel SINR and rate coverage probability expressions for DL and UL transmissions, respectively, incorporating the LOS probability model, beamforming gain per BS, cell-association bias and molecular absorption effect of the THz band.
Our numerical results reveal that a significant gain in both SINR and rate coverage probabilities can be achieved by raising the THz BS density.
It is noteworthy that the propagation loss and noise caused by molecular absorption in the THz band have an adverse impact on the SINR coverage probability. 
Moreover, we have demonstrated that the DL and UL decoupled cell-association strategy enables a more flexible bias factor design that ensures good SINR and rate coverage for both DL and UL communications.
In the future, we will extend the proposed framework to study the effects of beam misalignment and practical hardware imperfections. 

% Practical hardware imperfections will be incorporated into the system model.
% The performance evaluation for cell-edge UE in hybrid sub-6GHz-mmWave-THz networks is also of future interest.

\appendices
{

\section{Proof of Lemma \ref{lemma_sub6_AP}}
\label{App:sub6_AP}
Under the  strongest average biased
received signal cell-association strategy, the typical UE connects to the $s^{\mathrm{th}}$ sub-6GHz tier for the DL and UL if $P^{\mathrm{S},q}_{r,s}(R_{s}) > \mathop{\mathrm{max}}\limits_{g\in \mathcal{S},g\neq s}P^{\mathrm{S},q}_{r,g}(R_{g})$, $P^{\mathrm{S},q}_{r,s}(R_{s}) > \mathop{\mathrm{max}}\limits_{i\in \mathcal{M}}P^{\mathrm{M},q}_{r,i}(D_{i})$ and $P^{\mathrm{S},q}_{r,s}(R_{s}) > \mathop{\mathrm{max}}\limits_{j\in \mathcal{T}}P^{\mathrm{T},q}_{r,j}(D_{j})$, where {$q\in\{\mathrm{DL, UL}\}$.}
Denoting $R_{s}=x$, the association probability of a typical UE served by the nearest sub-6GHz BS in the $s^{\mathrm{th}}$ tier can be derived as
\begin{align}
\label{eq:sub6_AP}
\mathcal{A}^{q}_{s} = \mathbb{E}_x\Bigg[ 
&\mathbb{P}\left[ P^{\mathrm{S},q}_{r,s}(x) > \mathop{\mathrm{max}}\limits_{g\in \mathcal{S},g\neq s}P^{\mathrm{S},q}_{r,g}(R_{g}) \right] \times \nonumber\\
&\mathbb{P}\left[ P^{\mathrm{S},q}_{r,s}(x) > \mathop{\mathrm{max}}\limits_{i\in \mathcal{M}}P^{\mathrm{M},q}_{r,i}(D_{i}) \right] \times \nonumber\\
&\mathbb{P}\left[ P^{\mathrm{S},q}_{r,s}(x) > \mathop{\mathrm{max}}\limits_{j\in \mathcal{T}}P^{\mathrm{T},q}_{r,j}(D_{j}) \right]  \Bigg] ,
\end{align}
where  $q\in\{\mathrm{DL, UL}\}$.
Substituting (\ref{eq:sub6_RP}) (\ref{eq:mmWave_RP}) and (\ref{eq:Thz_RP}) into (\ref{eq:sub6_AP}), we have
\begin{align}
\label{eq:sub6_AP_1}
&\mathbb{P}\left[ P^{\mathrm{S},q}_{r,s}(x) > \mathop{\mathrm{max}}\limits_{g\in \mathcal{S},g\neq s}P^{\mathrm{S},q}_{r,g}(R_{g}) \right]  \nonumber\\
&= \prod_{g\in \mathcal{S},g\neq s}\mathbb{P}\left[P_{s}^{q}l^{\mathrm{S}}_{s}(x)C_{s}^{q} > P_{g}^{q}l^{\mathrm{S}}_{g}(R_{g})C_{g}^{q}\right]   \nonumber\\
&=  \prod_{g\in \mathcal{S},g\neq s}\mathbb{P}\Bigg[P_{s}^{q}\beta_{0} x^{-\alpha_{s}}C_{s}^{q} > P_{g}^{q}\beta_{0} R_{g}^{-\alpha_{g}}C_{g}^{q} \Bigg]    \nonumber\\
&=  \prod_{g\in \mathcal{S},g\neq s}\mathbb{P}\left[  R_{g} > \left(\frac{C_g^{q}P_g^{q}}{C_s^{q}P_s^{q}} \right)^{\frac{1}{\alpha_{g}}}x^{\frac{\alpha_{s}}{\alpha_{g}}} \right]     \nonumber\\
&\overset{(a)}{=} \prod_{g\in \mathcal{S},g\neq s} \mathrm{exp}\left( -2\pi\lambda_{g}\int_{0}^{\varrho_{s,g}^{q}(x)} u \mathrm{d}u \right)   \nonumber\\ 
&= \prod_{g\in \mathcal{S},g\neq s} \mathrm{exp}\left( -\pi\lambda_{g} \left(\varrho_{s,g}^{q}(x)\right)^2 \right),
\end{align}
\begin{align}
\label{eq:sub6_AP_2}
&\mathbb{P}\left[ P^{\mathrm{S},q}_{r,s}(x) > \mathop{\mathrm{max}}\limits_{i\in \mathcal{M}}P^{\mathrm{M},q}_{r,i}(D_{i}) \right] \nonumber\\
&= \prod_{i\in \mathcal{M} }\mathbb{P}\left[P_{s}^{q}l^{\mathrm{S}}_{s}(x)C_{s}^{q} > P_{i}^{q}G_{i}^{\mathrm{max}}l^{\mathrm{M}}_{i}(D_{i})C_{i}^{q}\right]   \nonumber\\
&=  \prod_{i\in \mathcal{M} }\mathbb{P}\left[  D_{i}>\left(\frac{C_i^{q}P_i^{q}N_i}{C_s^{q}P_s^{q}\beta_{0}}\left(\frac{c}{4\pi f_{\mathrm{M}}}\right)^{2} \right)^{\frac{1}{\alpha_{i}}}x^{\frac{\alpha_{s}}{\alpha_{i}}} \right]     \nonumber\\
&\overset{(b)}{=} \prod_{i\in \mathcal{M} } \mathrm{exp}\left( -2\pi\lambda_{i}\int_{0}^{\varepsilon_{s,i}^{q}(x)} P_{\mathrm{LOS}}(u)u \mathrm{d}u \right)   \nonumber\\ 
&= \prod_{i\in \mathcal{M}} \! \mathrm{exp}\!\left(\! \frac{2\pi\lambda_{i} e^{ -(\zeta \varepsilon_{s,i}^{q}(x) + p)}\!\!\left(\!1 \!-\! e^{\zeta \varepsilon_{s,i}^{q}(x)} \!+\! \zeta \varepsilon_{s,i}^{q}(x)\!\right)}{\zeta^2} \! \right) \!,
\end{align}
and
\begin{align}
\label{eq:sub6_AP_3}
&\mathbb{P}\left[ P^{\mathrm{S},q}_{r,s}(x) > \mathop{\mathrm{max}}\limits_{j\in \mathcal{T}}P^{\mathrm{T},q}_{r,j}(D_{j}) \right]   \nonumber\\
&=\prod_{j\in\mathcal{T}}\mathbb{P}\left[P_{s}^{q}l^{\mathrm{S}}_{s}(x)C_{s}^{q} > P_{j}^{q}G_{j}^{\mathrm{max}}l^{\mathrm{T}}_{j}(D_{j})C_{j}^{q}\right]   \nonumber\\
&=\prod_{j\in\mathcal{T}}\mathbb{P}\left[ D_{j}^{\alpha_{j}}e^{K_aD_{j}} > \left(\frac{C_j^{q}P_j^{q}N_j}{C_s^{q}P_s^{q}\beta_{0}}\left(\frac{c}{4\pi f_{\mathrm{T}}}\right)^{2}\right)x^{\alpha_{s}} \right]   \nonumber\\
&=\prod_{j\in\mathcal{T}}\mathbb{P}\left[D_{j}e^{\frac{K_aD_{j}}{\alpha_{j}}} > \left(\frac{C_j^{q}P_j^{q}N_j}{C_s^{q}P_s^{q}\beta_{0}}\left(\frac{c}{4\pi f_{\mathrm{T}}}\right)^{2}\right)^{\frac{1}{\alpha_{j}}}x^{\frac{\alpha_{s}}{\alpha_{j}}} \right]   \nonumber\\
&\overset{(c)}{=}\! \prod_{j\in\mathcal{T}}\mathbb{P}\Bigg[\frac{K_aD_{j}}{\alpha_{j}}e^{\frac{K_aD_{j}}{\alpha_{j}}} \!>\! \frac{K_a}{\alpha_{j}} \left(\frac{C_j^{q}P_j^{q}N_j}{C_s^{q}P_s^{q}\beta_{0}}\left(\frac{c}{4\pi f_{\mathrm{T}}}\right)^{2}\right)^{\frac{1}{\alpha_{j}}}\!\!x^{\frac{\alpha_{s}}{\alpha_{j}}} \Bigg]   \nonumber\\
&\overset{(d)}{=}\prod_{j\in\mathcal{T}}\mathbb{P}\left[\frac{K_aD_{j}}{\alpha_{j}}e^{\frac{K_aD_{j}}{\alpha_{j}}} > \Lambda_{s,j}^{\mathrm{S},q}x^{\frac{\alpha_{s}}{\alpha_{j}}} \right]   \nonumber\\
&\overset{(e)}{=} \prod_{j\in\mathcal{T}}\mathbb{P}\left[D_{j} > \frac{\alpha_{j}}{K_a}W_{\mathrm{L}}(\Lambda_{s,j}^{\mathrm{S},q}x^{\frac{\alpha_{s}}{\alpha_{j}}}) \right]    \nonumber\\
&\overset{(f)}{=} \prod_{j\in \mathcal{T}} \mathrm{exp}\left( -2\pi\lambda_{j}\int_{0}^{\vartheta_{s,j}^{q}(x)} P_{\mathrm{LOS}}(u)u \mathrm{d}u \right)  \nonumber\\ 
&= \prod_{j\in \mathcal{T}} \mathrm{exp}\left(\! \frac{2\pi\lambda_{j} e^{ -(\zeta \vartheta_{s,j}^{q}(x) + p)}\!\!\left(1 \!-\! e^{\zeta \vartheta_{s,j}^{q}(x) } \!+\! \zeta \vartheta_{s,j}^{q}(x)\!\right)}{\zeta^2} \! \right) \!,
\end{align}
where (a), (b), and (f) come from the null probability of HPPP with $\varrho_{s,g}^{q}(x)=\left(\frac{C_g^{q}P_g^{q}}{C_s^{q}P_s^{q}}\right)^{\frac{1}{\alpha_{g}}}x^{\frac{\alpha_{s}}{\alpha_{g}}}$,  $\varepsilon_{s,i}^{q}(x)=\left(\frac{C_i^{q}P_i^{q}N_i}{C_s^{q}P_s^{q}\beta_{0}}\left(\frac{c}{4\pi f_{\mathrm{M}}}\right)^{2}\right)^{\frac{1}{\alpha_{i}}}x^{\frac{\alpha_{s}}{\alpha_{i}}}$, and $\vartheta_{s,j}^{q}(x)=\frac{\alpha_{j}}{K_a}W_{\mathrm{L}}\left(\Lambda_{s,j}^{\mathrm{S},q}x^{\frac{\alpha_{s}}{\alpha_{j}}}\right)$,  (c) is to multiply $\frac{K_a}{\alpha_{j}}$ on both sides of the equation, 
$\Lambda_{s,j}^{\mathrm{S},q}=\frac{K_a}{\alpha_{j}}\left(\frac{C_j^{q}P_j^{q}N_j}{C_s^{q}P_s^{q}\beta_{0}}\left(\frac{c}{4\pi f_{\mathrm{M}}}\right)^{2} \right)^{\frac{1}{\alpha_j}}$ in (d), and
$W_{\mathrm{L}}(\cdot)$ in (e) is the Lambert W function, which is given by $x=W_{\mathrm{L}}(x)e^{W_{\mathrm{L}}(x)}$.

\section{Proof of Lemma \ref{lemma_sub6_PDF}}
\label{App:sub6_PDF}
When the typical UE is associated with a BS in the $s^{\mathrm{th}}$ sub-6 GHz tier, the probability of the event $X_{s}^{q}<x$ can be derived as
\begin{align}
\mathbb{P}\left[X_{s}^{q}<x\right] 
= \mathbb{P}\left[X_{s}^{q}<x|v^{q}=s \right] 
=\frac{\mathbb{P}\left[R_{s}<x, v^{q}=s \right]}{\mathbb{P}\left[v^{q}=s \right]}  ,
\end{align}
where $q\in\{\mathrm{DL, UL}\}$, $v^{q}$ is the index of the layer that the typical UE associated with and $\mathbb{P}\left[v^{q}=s \right]=\mathcal{A}_{s}^{q}$. The joint probability of $\mathbb{P}\left[R_{s}<x, v^{q}=s \right]$ can be derived as
\begin{align}
\label{eq:mmWave_f_Xm_1}
&\mathbb{P}\left[R_{s}<x, v^{q}=s \right] \nonumber\\
&= \int_{0}^{x} \mathbb{P}\left[ P^{\mathrm{S},q}_{r,s}(r) > \mathop{\mathrm{max}}\limits_{g\in \mathcal{S},g\neq s}P^{\mathrm{S},q}_{r,g}(R_{g}) \right] \times \nonumber\\
&\qquad\quad \mathbb{P}\left[ P^{\mathrm{S},q}_{r,s}(r) > \mathop{\mathrm{max}}\limits_{i\in \mathcal{M}}P^{\mathrm{M},q}_{r,i}(D_{i}) \right] \times \nonumber\\
&\qquad\quad \mathbb{P}\left[ P^{\mathrm{S},q}_{r,s}(r) > \mathop{\mathrm{max}}\limits_{j\in \mathcal{T}}P^{\mathrm{T},q}_{r,j}(D_{j}) \right] f_{R_{s}}(r) \mathrm{d}r  ,
\end{align}
where the expressions of $\mathbb{P}\left[ P^{\mathrm{S},q}_{r,s}(r) > \mathop{\mathrm{max}}\limits_{g\in \mathcal{S},g\neq s}P^{\mathrm{S},q}_{r,g}(R_{g}) \right]$, $\mathbb{P}\left[\! P^{\mathrm{S},q}_{r,s}(r) \!\!>\!\! \mathop{\mathrm{max}}\limits_{i\in \mathcal{M}}P^{\mathrm{M},q}_{r,i}(D_{i}) \!\right]$, and $\mathbb{P}\left[\! P^{\mathrm{S},q}_{r,s}(r) \!\!>\! \!\mathop{\mathrm{max}}\limits_{j\in \mathcal{T}}P^{\mathrm{T},q}_{r,j}(D_{j}) \!\right]$ can be found in (\ref{eq:sub6_AP_1}), (\ref{eq:sub6_AP_2}), and (\ref{eq:sub6_AP_3}) in Appendix \ref{App:sub6_AP}, respectively. 
Recall that we only consider NLOS transmission links for sub-6 GHz tiers and $R_{s}$ is the distance from the typical UE to its nearest BS in the $s^{\mathrm{th}}$ tier.
The PDF of $R_{s}$ is given by $f_{R_{s}}(r)=2\pi\lambda_s re^{-\pi\lambda_s r^2}$ \cite{Andrews2011Tractable}.
Then the PDF of $X_{s}^{q}$ can be obtained as
\begin{align}
f_{X_{s}^{q}}(x)&=\frac{\mathrm{d}\mathbb{P}\left[X_{s}^{q}<x\right]}{\mathrm{d}x}.
\end{align}
}

{
\section{Proof of Theorem \ref{theorem_sub6_CCP}}
\label{App:sub6_CCP}
When the typical UE connects to the $s^{\mathrm{th}}$ sub-6 GHz tier, the conditional coverage probability is computed by
\begin{align}
&\mathbb{P}\left[\mathrm{SINR}_{s}^{q}(x)>\tau\right]  = \mathbb{P}\left[ \frac{P_{s}^{q}l^{\mathrm{S}}_{s}(x)}{I_{s}^{\mathrm{S}, q} + \delta_{s}^2} > \tau \right]   \nonumber\\
&\approx \mathbb{P}\left[ \frac{P_{s}^{q}\beta_{0}h_{s}x^{-\alpha_{s}}}{\sum\limits_{j\in \mathcal{S}} \sum\limits_{i\in \Phi_{j}\backslash B_{s}^{q}} P_{j}^{q}\beta_{0}h_{j}d_{i,j}^{-\alpha_{j}} + \delta_{s}^2}  > \tau \right]   \nonumber\\
&= \mathbb{P}\Bigg[h_{s} > \frac{\tau x^{\alpha_{s}}}{P_{s}^{q}}  \Bigg(\sum\limits_{j\in \mathcal{S}} \sum\limits_{i\in \Phi_{j}\backslash B_{s}^{q}} P_{j}^{q}h_{j}d_{i,j}^{-\alpha_{j}} + \frac{\delta_{s}^2}{\beta_{0}} \Bigg)\Bigg]   \nonumber\\
&= \mathbb{E}\!\left[\mathrm{exp}\left( -\frac{\tau x^{\alpha_{s}}}{P_{s}^{q}}  \left(\sum\limits_{j\in \mathcal{S}} \sum\limits_{i\in \Phi_{j}\backslash B_{s}^{q}} P_{j}^{q}h_{j}d_{i,j}^{-\alpha_{j}} + \frac{\delta_{s}^2}{\beta_{0}} \right)\right)\right]  \nonumber\\
&=\prod\limits_{j\in \mathcal{S}}\mathbb{E}\!\left[ \mathrm{exp}\!\!\left( -Y_s^{q}(\tau) \sum\limits_{i\in \Phi_{j}\backslash B_{s}^{q}} \!\!\!\! P_{j}^{q}h_{j}d_{i,j}^{-\alpha_{j}} \!\! \right) \!\! \right] \mathrm{exp}\!\left( -Y_s^{q}(\tau)O^{\mathrm{S}}_s\right)   \nonumber\\
&=\prod\limits_{j\in \mathcal{S}}\mathbb{E}\!\left[\prod\limits_{i\in \Phi_{j}\backslash B_{s}^{q}} \frac{d_{i,j}^{\alpha_{j}}}{d_{i,j}^{\alpha_{j}} + Y_s^{q}(\tau)P_{j}^{q}} \right] \mathrm{exp}\!\left( -Y_s^{q}(\tau)O^{\mathrm{S}}_s\right)  \nonumber\\
&\overset{(a)}{=}\prod\limits_{j\in \mathcal{S}}\mathrm{exp}\left(-2\pi\lambda_j \int_{\varrho_{s,j}^{q}(x)}^{\infty}\left(1- \frac{d_{i,j}^{\alpha_{j}}}{d_{i,j}^{\alpha_{j}} + Y_s^{q}(\tau)P_{j}^{q}}\right)r \mathrm{d}r\right)  \nonumber\\ &\qquad\qquad\qquad\qquad\qquad\qquad\ \ 
\times\mathrm{exp}\!\left( -Y_s^{q}(\tau)O^{\mathrm{S}}_s\right),
\end{align}
where $O^{\mathrm{S}}_s=\frac{\delta_{s}^2}{\beta_{0}}$, $Y_s^{q}(\tau) = \frac{\tau x^{\alpha_{s}}}{P_{s}^{q}}$, and $\varrho_{s,j}^{q}(x)=\left(\frac{C_j^{q}P_j^{q}}{C_s^{q}P_s^{q}} \right)^{\frac{1}{\alpha_{j}}}x^{\frac{\alpha_{s}}{\alpha_{j}}}$. (a) applies
the probability generating function of HPPP. In particular, for a HPPP $\Phi$ with intensity $\lambda$, we have $\mathbb{E}\left[ \prod_{x\in\Phi} f(x)\right] = \mathrm{exp}\left( -\lambda\int_{\mathbb{R}^2}(1-f(x))\mathrm{d}x\right)$ \cite{Andrews2011Tractable}.
}

\section{Proof of Theorem \ref{theorem_mmWave_CCP}}
\label{App:mmWave_CCP}
When the typical UE connects to the $m^{\mathrm{th}}$ mmWave tier, the conditional coverage probability is computed by
\begin{align}
&\mathbb{P}\left[\mathrm{SINR}_{m}^{q}(x)>\tau\right]  = \mathbb{P}\left[ \frac{P_{m}^{q}G_{m}^{\mathrm{max}}l^{\mathrm{M}}_{m}(x)}{I_{m}^{\mathrm{M}, q} + \delta_{m}^2} > \tau \right]   \nonumber\\
&\overset{(a)}{\approx} \mathbb{P}\left[ \frac{P_{m}^{q}N_{m}\left(\frac{c}{4\pi f_{\mathrm{M}}}\right)^{2}h_{m}x^{-\alpha_{m}}}{\sum\limits_{j\in \mathcal{M}} \sum\limits_{i\in \Phi_{j}^{\mathrm{L}}\backslash B_{m}^{q}} P_{j}^{q}G_{j}\left(\frac{c}{4\pi f_{\mathrm{M}}}\right)^{2}h_{j}d_{i,j}^{-\alpha_{j}} + \delta_{m}^2}  > \tau \right]   \nonumber\\
&= \mathbb{P}\Bigg[h_{m} > \frac{\tau x^{\alpha_{m}}}{P_{m}^{q}N_{m}} \times  \nonumber\\
&\quad\quad\quad\quad 
\Bigg( \sum\limits_{j\in \mathcal{M}} \sum\limits_{i\in \Phi_{j}^{\mathrm{L}}\backslash B_{m}^{q}} \!P_{j}^{q}G_{j}h_{j}d_{i,j}^{-\alpha_{j}}  +\delta_{m}^{2}\left(\frac{4\pi f_{\mathrm{M}}}{c}\right)^{2} \Bigg) \Bigg]   \nonumber\\
&\overset{(b)}{<} 1 - \mathbb{E}\left[ \left(1  \!-\!  \mathrm{exp}\Bigg( \!-\frac{\eta_{m}\tau x^{\alpha_{m}}}{P_{m}^{q}N_{m}} \bigg(\sum\limits_{j\in \mathcal{M}}  H_j^{q}  + O^{\mathrm{M}}_m  \bigg) \Bigg) \right)^{\!\gamma_m} \right]    \nonumber\\
&\overset{(c)}{=} \sum_{n=1}^{\gamma_m}(-1)^{n+1}\left(\substack{\gamma_m \\ \\ n}\right)\times  \nonumber \\
 &\quad \ \ \quad\quad\quad\quad\mathbb{E}\left[\mathrm{exp}\left( - \frac{n\eta_{m}\tau x^{\alpha_{m}}}{P_{m}^{q}N_{m}} \bigg( \sum\limits_{j\in \mathcal{M}}  H_j^{q}  +  O^{\mathrm{M}}_m  \bigg) \right) \right]   \nonumber\\
&= \sum_{n=1}^{\gamma_m}( - 1)^{n + 1}\left( \substack{\gamma_m \\ \\ n}\right) \times \nonumber \\
& \quad\quad\quad\quad\mathbb{E}\left[\mathrm{exp}\left( -V_m^{q}(\tau) \sum\limits_{j\in \mathcal{M}}H_j^{q} \right) \right] \mathrm{exp}\left( - V_m^{q}(\tau)O^{\mathrm{M}}_m\right)  \nonumber\\
&= \sum_{n=1}^{\gamma_m}( - 1)^{n + 1}\left( \substack{\gamma_m \\ \\ n}\right) \times \nonumber\\
&\quad\quad\quad\prod\limits_{j\in \mathcal{M}} \mathbb{E}\left[\mathrm{exp}\left(  - V_m^{q}(\tau)H_j^{q} \right)\right] \mathrm{exp}\left( -V_m^{q}(\tau)O^{\mathrm{M}}_m\right), 
\end{align}
where we rewrite $G_{j}(N_j, \phi_{{D}_{i, j}})$ as $G_j$ in (a) for clarity, $H_j^{q}= \sum\limits_{i\in \Phi_{j}^{\mathrm{L}}\backslash B_{m}^{q}} P_{j}^{q}G_{j}h_{j}d_{i,j}^{-\alpha_{j}}$, $O^{\mathrm{M}}_m=\delta_{m}^{2}\left(\frac{4\pi f_{\mathrm{M}}}{c}\right)^{2}$, and $V_m^{q}(\tau) = \frac{n\eta_{m}\tau x^{\alpha_{m}}}{P_{m}^{q}N_{m}}$. {(b) is because for a gamma random variable $h_{m}$ with shape parameter $\gamma_m$, i.e., $h_{m}\sim\Gamma(\gamma_m,\frac{1}{\gamma_m})$, the probability $\mathbb{P}\left( h_{m} > x\right)$ can be tightly upper-bounded by $\mathbb{P}\left( h_{m} > x\right)<1 - \left( 1-e^{-\eta_{m} x} \right)^{\gamma_m} $, where $\eta_{m}=\gamma_m(\gamma_m !)^{-\frac{1}{\gamma_m}}$ \cite{alzer1997some, thornburg2016performance}.} (c) is derived following Binomial series expansion. Then, the Laplace transform of $H_j^{q}$ can be obtained as follows
\begin{align}
&\mathbb{E}\left[ \mathrm{exp}\left( -V_m^{q}(\tau)H_j^{q} \right)\right] \nonumber\\
&=\mathbb{E}\left[ \mathrm{exp}\left( -V_m^{q}(\tau)\sum\limits_{i\in \Phi_{j}^{\mathrm{L}}\backslash B_{m}^{q}} P_{j}^{q}G_{j}h_{j}d_{i,j}^{-\alpha_{j}} \right)\right]
\nonumber\\
&\overset{(d)}{=} \!{\mathrm{exp}\!\Bigg(\!\!-\!2\pi\lambda_j\!\!\!\!\!\!\!\!\sum_{w\in \{\mathrm{max},\mathrm{min}\}}\!\!\!\!\!\!\!\!P_{\mathrm{G},j}^w \!\!\int_{\Omega_{m,j}^{q}}^{\infty}\!\!\!\!\!\! \left(\! 1\!-\!\mathbb{E}_{h_{j}}\!\!\left[ e^{-V_m^{q}(\tau)P_{j}^{q}G_{j}^wh_j r^{-\alpha_{j}}} \right] \! \right)}  \nonumber\\ 
&\qquad\qquad\qquad\qquad\qquad\qquad\quad  {\times P_{\mathrm{LOS}}(r)r \mathrm{d}r \Bigg)}    \nonumber\\
&= \!\!\!\! \prod\limits_{w\in \{\mathrm{max},\mathrm{min}\}} \!\!\!\!\!\!\!\! \mathrm{exp}\Bigg( \!-\!2\pi\lambda_jP_{\mathrm{G},j}^w \!\!\int_{\Omega_{m,j}^{q}}^{\infty}\!\!\!\!\!\! \left( 1\!-\!\mathbb{E}_{h_{j}}\!\!\left[ {e^{-V_m^{q}(\tau)P_{j}^{q}G_{j}^wh_{j} r^{-\alpha_{j}}}} \right]  \right)  \nonumber\\ 
&\qquad\qquad\qquad\qquad\qquad\qquad\quad \times P_{\mathrm{LOS}}(r)r \mathrm{d}r \Bigg)    \nonumber\\
&\overset{(e)}{=} \!\!\!\! \prod\limits_{w\in \{\mathrm{max},\mathrm{min}\}} \!\!\!\!\!\! \mathrm{exp}\Bigg( -2\pi\lambda_jP_{\mathrm{G},j}^w \int_{\Omega_{m,j}^{q}}^{\infty} P_{\mathrm{LOS}}(r) r  \times \nonumber\\
&\qquad\qquad\qquad\quad \Bigg( 1 - \left( 1 + \frac{V_m^{q}(\tau)P_{j}^{q}G_{j}^w r^{-\alpha_{j}}}{\gamma_j}  \right)^{-\gamma_j} \Bigg) \mathrm{d}r \Bigg)   \nonumber\\
&\overset{(f)}{=} \!\!\!\! \prod\limits_{w\in \{\mathrm{max},\mathrm{min}\}} \!\!\!\!\!\!\!\!\mathrm{exp}\left( -2\pi\lambda_jP_{\mathrm{G},j}^w \int_{\Omega_{m,j}^{q}}^{\infty} \!\!\!\!\!\!\left( 1 \!-\! \Delta_{m,j}^{b,w}(r) \right) P_{\mathrm{LOS}}(r) r \mathrm{d}r \right).
\end{align}
In (d), we have $\Omega_{m,j}^{q}=\left(\frac{C_j^{q}P_j^{q}N_j}{C_m^{q}P_m^{q}N_m} \right)^{\frac{1}{\alpha_{j}}}x^{\frac{\alpha_{m}}{\alpha_{j}}}$, $P_{\mathrm{G},j}^{\mathrm{max}}=\phi_{\mathrm{3dB},j}/0.5=2\phi_{\mathrm{3dB},j}$ is the probability that the typical UE (BS) is located in the main-lobe direction, and $P_{\mathrm{G},j}^{\mathrm{min}}=1-P_{\mathrm{G},j}^{\mathrm{max}}$ is the probability that the typical UE (BS) is located in the side-lobe direction. (e) follows the moment generating function of the gamma random variable {$h_{j}\sim\Gamma(\gamma_j,\frac{1}{\gamma_j})$, which is given by $\mathbb{E}_{h_j}[e^{h_{j}x}]=\left( 1 - \frac{x}{\gamma_j}  \right)^{-\gamma_j}$}.
In (f), $\Delta_{m,j}^{b,w}(r)= \left( 1 + \frac{V_m^{q}(\tau)P_{j}^{q}G_{j}^w r^{-\alpha_{j}}}{\gamma_j}  \right)^{-\gamma_j}$ and the lower bound of the integral is the minimum distance between the interfering BS (UE) in the $j^{\mathrm{th}}$ tier and the typical UE (BS).

\section{Proof of Theorem \ref{theorem_Thz_CCP}}
\label{App:Thz_CCP}
When the typical UE is connected to the $t^{\mathrm{th}}$ THz tier, the conditional coverage probability is computed by
\begin{align}
&\mathbb{P}[\mathrm{SINR}_{t}^{q}(x)>\tau] = \mathbb{P}\left[ \frac{P_{t}^{q}G_{t}^{\mathrm{max}}\left(\frac{c}{4\pi f_{\mathrm{T}}}\right)^{2}x^{-\alpha_{t}}e^{-K_ax}}{I_{t}^{\mathrm{T},q} + \mathrm{Noise}_{t}^{q}(x)} > \tau \right]   \nonumber\\
&\overset{(a)}{\approx} \mathbb{P}\left[ \frac{P_{t}^{q}G_{t}^{\mathrm{max}}\left(\frac{c}{4\pi f_{\mathrm{T}}}\right)^{2}x^{-\alpha_{t}}e^{-K_ax}h_{\mathrm{T}}}{I_{t}^{\mathrm{T},q} + \mathrm{Noise}_{t}^{q}(x)} > \tau \right]   \nonumber\\
&\overset{(b)}{\approx} \mathbb{P}\Bigg[J_t^{q}(x)e^{-K_ax}h_{\mathrm{T}}\Bigg(J_t^{q}(x)\left(1-e^{-K_ax}\right) + \delta_{t}^2 +  \nonumber\\
&\qquad\qquad\qquad  \sum\limits_{j\in  \mathcal{T}} \sum\limits_{i\in \Phi_{j}^{\mathrm{L}}\backslash B_{t}^{q}}P_{j}^{q}G_{j}\left(\frac{c}{4\pi f_{\mathrm{T}}}\right)^{2}d_{i,j}^{-\alpha_{t}} \Bigg)^{-1} > \tau \Bigg]   \nonumber\\
&= \mathbb{P}\Bigg[h_{\mathrm{T}} > \frac{\tau e^{K_ax}}{J_t^{q}(x)} \Bigg(J_t^{q}(x)\left(1-e^{-K_ax}\right) + \delta_{t}^2 +  \nonumber\\
&\qquad\qquad\qquad\qquad\quad \sum\limits_{j\in  \mathcal{T}} \sum\limits_{i\in \Phi_{j}^{\mathrm{L}}\backslash B_{t}^{q}}P_{j}^{q}G_{j}\left(\frac{c}{4\pi f_{\mathrm{T}}}\right)^{2}d_{i,j}^{-\alpha_{t}} \Bigg)\Bigg],
\end{align}
where in (a), we induce a gamma random variable $h_{\mathrm{T}}\sim\Gamma(\gamma_\mathrm{T},\frac{1}{\gamma_\mathrm{T}})$ to facilitate further derivations, where $\gamma_\mathrm{T}$ is the shape parameter and when $\gamma_{\mathrm{T}}\rightarrow \infty$, $h_{\mathrm{T}}\rightarrow 1$. In (b), we rewrite $G_{j}(N_j, \phi_{{D}_{i, j}})$ as $G_j$ for clarity and $J_t^{q}(x)=P_{t}^{q}G_{t}^{\mathrm{max}}\left(\frac{c}{4\pi f_{\mathrm{T}}}\right)^{2}x^{-\alpha_{t}}$.

Denoting $Q_t^{q}=\sum\limits_{j\in  \mathcal{T}} \sum\limits_{i\in \Phi_{j}^{\mathrm{L}}\backslash B_{t}^{q}}P_{j}^{q}G_{j}\left(\frac{c}{4\pi f_{\mathrm{T}}}\right)^{2}d_{i,j}^{-\alpha_{t}}$, we have
\begin{align}
&\mathbb{P}\Bigg[h_{\mathrm{T}} > \frac{\tau e^{K_ax}}{J_t^{q}(x)} \Bigg(J_t^{q}(x)\left(1-e^{-K_ax}\right) + \delta_{t}^2 + Q_t^{q} \Bigg) \Bigg]   \nonumber\\
&< 1 - \mathbb{E}\Bigg[\Bigg(1 -  \mathrm{exp}\bigg( -\eta_{\mathrm{T}}\frac{\tau e^{K_ax}}{J_t^{q}(x)} \times \nonumber\\ 
&\qquad\qquad\quad\ \  \left(J_t^{q}(x)\left(1-e^{-K_ax}\right) + \delta_{t}^2 + Q_t^{q} \right) \bigg) \Bigg)^{\gamma_{\mathrm{T}}} \Bigg]   \nonumber\\
&= \sum_{n=1}^{\gamma_{\mathrm{T}}}(-1)^{n+1}\left( \substack{\gamma_{\mathrm{T}} \\ \\ n}\right) \mathbb{E}\Bigg[\mathrm{exp}\Bigg( - n\eta_{\mathrm{T}}\frac{\tau e^{K_ax}}{J_t^{q}(x)}\times \nonumber\\ 
&\qquad\qquad\qquad\quad\,\, \bigg(J_t^{q}(x)\left(1-e^{-K_ax}\right) + \delta_{t}^2 + Q_t^{q} \bigg) \Bigg) \Bigg]   \nonumber\\
&= \sum_{n=1}^{\gamma_{\mathrm{T}}}(-1)^{n+1}\left( \substack{\gamma_{\mathrm{T}} \\ \\ n}\right) \mathbb{E}\Bigg[\mathrm{exp}\Bigg( - n\eta_{\mathrm{T}}\frac{\tau e^{K_ax}}{J_t^{q}(x)}Q_t^{q} \Bigg)\Bigg] \times \nonumber\\ 
&\qquad\qquad\ \   \mathrm{exp}\Bigg( - n\eta_{\mathrm{T}}\tau \bigg(\frac{\delta_{t}^2 e^{K_ax}}{J_t^{q}(x)} + e^{K_ax} - 1 \bigg)\Bigg) ,  
\end{align}
where $\eta_{\mathrm{T}}=\gamma_{\mathrm{T}}(\gamma_{\mathrm{T}} !)^{-\frac{1}{\gamma_{\mathrm{T}}}}$, and the Laplace transform of $Q_t^{q}$ can be computed as
\begin{align}
&\mathbb{E}\Bigg[\mathrm{exp}\Bigg( - n\eta_{\mathrm{T}}\frac{\tau e^{K_ax}}{J_t^{q}(x)}Q_t^{q} \Bigg)\Bigg]  \nonumber\\ 
&= \mathbb{E}\Bigg[\mathrm{exp}\Bigg( - n\eta_{\mathrm{T}}\frac{\tau e^{K_ax}}{P_{t}^{q}G_{t}^{\mathrm{max}}\left(\frac{c}{4\pi f_{\mathrm{T}}}\right)^{2}x^{-\alpha_{t}}} \times  \nonumber\\ 
&\qquad\qquad\quad\quad \sum\limits_{j\in  \mathcal{T}} \sum\limits_{i\in \Phi_{j}^{\mathrm{L}}\backslash B_{t}^{q}}P_{j}^{q}G_{j}\left(\frac{c}{4\pi f_{\mathrm{T}}}\right)^{2}d_{i,j}^{-\alpha_{t}}\Bigg)\Bigg]  \nonumber\\
&=\prod\limits_{j\in  \mathcal{T}} \mathbb{E}\Bigg[\mathrm{exp}\Bigg(\!\! - n\eta_{\mathrm{T}}\tau x^{\alpha_{t}}e^{K_ax} \!\!\!\!\!\! \sum\limits_{i\in \Phi_{j}^{\mathrm{L}}\backslash B_{t}^{q}} \!\!\!\! \frac{P_{j}^{q}G_{j}}{P_{t}^{q}G_{t}^{\mathrm{max}}}d_{i,j}^{-\alpha_{t}} \!\!\Bigg)\!\Bigg]  \nonumber\\
&=\prod\limits_{j\in  \mathcal{T}} \mathrm{exp}\Bigg(	 -2\pi\lambda_j\sum_{w\in \{\mathrm{max},\mathrm{min}\}} P_{\mathrm{G},j}^w \int_{\Theta_{t,j}^{q}}^{\infty} P_{\mathrm{LOS}}(r)r \times \nonumber\\
&\qquad\quad \Bigg( 1- \mathrm{exp}\bigg( - n\eta_{\mathrm{T}}\tau x^{\alpha_{t}}e^{K_ax}  \frac{P_{j}^{q}G_{j}^{w}}{P_{t}^{q}N_{t}}r^{-\alpha_{t}}\bigg)\Bigg) \mathrm{d}r \Bigg) \nonumber\\
&=\prod\limits_{j\in  \mathcal{T}} \prod\limits_{w\in \{\mathrm{max},\mathrm{min}\}} \mathrm{exp}\Bigg(	 -2\pi\lambda_jP_{\mathrm{G},j}^w \int_{\Theta_{t,j}^{q}}^{\infty} P_{\mathrm{LOS}}(r)r \times \nonumber\\
&\qquad\Bigg( 1- \mathrm{exp}\bigg(-n\eta_{\mathrm{T}}\tau x^{\alpha_{t}}e^{K_ax}  \frac{P_{j}^{q}G_{j}^{w}}{P_{t}^{q}N_{t}}r^{-\alpha_{t}}\bigg)\Bigg) \mathrm{d}r \Bigg).  
\end{align}

\ifCLASSOPTIONcaptionsoff
  \newpage
\fi

\normalem


\begin{thebibliography}{10}

\bibitem{shokri2015millimeter}
H.~Shokri-Ghadikolaei, C.~Fischione, G.~Fodor, P.~Popovski, and M.~Zorzi, ``Millimeter wave cellular networks: A {MAC} layer perspective,'' {\em IEEE Trans. Commun.}, vol.~63, no.~10, pp.~3437--3458, Oct. 2015.

\bibitem{wang2023road}
C.-X. Wang, X.~You, X.~Gao, X.~Zhu, Z.~Li, C.~Zhang, H.~Wang, Y.~Huang, Y.~Chen, H.~Haas, {\em et~al.}, ``On the road to {6G}: Visions, requirements, key technologies and testbeds,'' {\em IEEE Commun. Surv. Tutor.}, 2023.

\bibitem{rappaport2019wireless}
T.~S. Rappaport, Y.~Xing, O.~Kanhere, S.~Ju, A.~Madanayake, S.~Mandal, A.~Alkhateeb, and G.~C. Trichopoulos, ``Wireless communications and applications above 100 {GHz}: Opportunities and challenges for {6G} and beyond,'' {\em IEEE Access}, vol.~7, pp.~78729--78757, 2019.

\bibitem{han2022terahertz}
C.~Han, Y.~Wang, Y.~Li, Y.~Chen, N.~A. Abbasi, T.~K{\"u}rner, and A.~F. Molisch, ``Terahertz wireless channels: A holistic survey on measurement, modeling, and analysis,'' {\em IEEE Commun. Surv. Tutor.}, vol.~24, no.~3, pp.~1670--1707, 2022.

\bibitem{rangan2014millimeter}
S.~Rangan, T.~S. Rappaport, and E.~Erkip, ``Millimeter-wave cellular wireless networks: Potentials and challenges,'' {\em Proc. IEEE}, vol.~102, no.~3, pp.~366--385, 2014.

\bibitem{deng2018millimeter}
N.~Deng, M.~Haenggi, and Y.~Sun, ``Millimeter-wave device-to-device networks with heterogeneous antenna arrays,'' {\em IEEE Trans. Commun.}, vol.~66, no.~9, pp.~4271--4285, Sept. 2018.

\bibitem{bai2014coverage}
T.~Bai and R.~W. Heath, ``Coverage and rate analysis for millimeter-wave cellular networks,'' {\em IEEE Trans. Wirel. Commun.}, vol.~14, no.~2, pp.~1100--1114, Feb. 2015.

\bibitem{turgut2017coverage}
E.~Turgut and M.~C. Gursoy, ``Coverage in heterogeneous downlink millimeter wave cellular networks,'' {\em IEEE Trans. Commun.}, vol.~65, no.~10, pp.~4463--4477, Oct. 2017.

\bibitem{Chen2021Optimal}
C.~Chen, J.~Zhang, X.~Chu, and J.~Zhang, ``On the optimal base-station height in {mmWave} small-cell networks considering cylindrical blockage effects,'' {\em IEEE Trans. Veh. Technol.}, vol.~70, no.~9, pp.~9588--9592, Sept. 2021.

\bibitem{chen2021coverage}
W.~Chen, L.~Li, Z.~Chen, and T.~Q. Quek, ``Coverage modeling and analysis for outdoor {THz} networks with blockage and molecular absorption,'' {\em IEEE Wirel. Commun. Lett.}, vol.~10, no.~5, pp.~1028--1031, May. 2021.

\bibitem{wu2020interference}
Y.~Wu, J.~Kokkoniemi, C.~Han, and M.~Juntti, ``Interference and coverage analysis for terahertz networks with indoor blockage effects and line-of-sight access point association,'' {\em IEEE Trans. Wirel. Commun.}, vol.~20, no.~3, pp.~1472--1486, Mar. 2021.

\bibitem{sayehvand2020interference}
J.~Sayehvand and H.~Tabassum, ``Interference and coverage analysis in coexisting {RF} and dense terahertz wireless networks,'' {\em IEEE Wirel. Commun. Lett.}, vol.~9, no.~10, pp.~1738--1742, Oct. 2020.

\bibitem{humadi2022user}
K.~Humadi, I.~Trigui, W.-P. Zhu, and W.~Ajib, ``User-centric cluster design and analysis for hybrid sub-6{GHz}-{mmWave-THz} dense networks,'' {\em IEEE Trans. Veh. Technol.}, vol.~71, no.~7, pp.~7585--7598, Jul. 2022.

\bibitem{sopin2022user}
E.~Sopin, D.~Moltchanov, A.~Daraseliya, Y.~Koucheryavy, and Y.~Gaidamaka, ``User association and multi-connectivity strategies in joint terahertz and millimeter wave {6G} systems,'' {\em IEEE Trans. Veh. Technol.}, vol.~71, no.~12, pp.~12765--12781, Dec. 2022.

\bibitem{hossan2021mobility}
M.~T. Hossan and H.~Tabassum, ``Mobility-aware performance in hybrid {RF} and terahertz wireless networks,'' {\em IEEE Trans. Commun.}, vol.~70, no.~2, pp.~1376--1390, Feb. 2022.

\bibitem{boccardi2016decouple}
F.~Boccardi, J.~Andrews, H.~Elshaer, M.~Dohler, S.~Parkvall, P.~Popovski, and S.~Singh, ``Why to decouple the uplink and downlink in cellular networks and how to do it,'' {\em IEEE Commun. Mag.}, vol.~54, no.~3, pp.~110--117, Mar. 2016.

\bibitem{lahad2020joint}
B.~Lahad, M.~Ibrahim, S.~Lahoud, K.~Khawam, and S.~Martin, ``Joint modeling of {TDD} and decoupled uplink/downlink access in {5G HetNets} with multiple small cells deployment,'' {\em IEEE Trans. Mob. Comput.}, vol.~20, no.~7, pp.~2395--2411, Jul. 2021.

\bibitem{dai2020decoupled}
C.~Dai, K.~Zhu, C.~Yi, and E.~Hossain, ``Decoupled uplink-downlink association in full-duplex cellular networks: A contract-theory approach,'' {\em IEEE Trans. Mob. Comput.}, vol.~21, no.~3, pp.~911--925, Mar. 2022.

\bibitem{Elshaer2016Downlink}
H.~Elshaer, M.~N. Kulkarni, F.~Boccardi, J.~G. Andrews, and M.~Dohler, ``Downlink and uplink cell association with traditional macrocells and millimeter wave small cells,'' {\em IEEE Trans. Wirel. Commun.}, vol.~15, no.~9, pp.~6244--6258, Sept. 2016.

\bibitem{wang2023performance}
Y.~Wang, C.~Chen, H.~Zheng, and X.~Chu, ``Performance of indoor small-cell networks under interior wall penetration losses,'' {\em IEEE Internet Things J.}, vol.~10, no.~12, pp.~10907--10915, Jun. 2023.

\bibitem{ding2017uplink}
T.~Ding, M.~Ding, G.~Mao, Z.~Lin, D.~L{\'o}pez-P{\'e}rez, and A.~Y. Zomaya, ``Uplink performance analysis of dense cellular networks with {LoS} and {NLoS} transmissions,'' {\em IEEE Trans. Wirel. Commun.}, vol.~16, no.~4, pp.~2601--2613, Apr. 2017.

\bibitem{bai2012using}
T.~Bai, R.~Vaze, and R.~W. Heath, ``Using random shape theory to model blockage in random cellular networks,'' in {\em SPCOM}, pp.~1--5, 2012.

\bibitem{bai2014analysis}
T.~Bai, R.~Vaze, and R.~W. Heath, ``Analysis of blockage effects on urban cellular networks,'' {\em IEEE Trans. Wirel. Commun.}, vol.~13, no.~9, pp.~5070--5083, Sept. 2014.

\bibitem{Lee2016Randomly}
G.~Lee, Y.~Sung, and J.~Seo, ``Randomly-directional beamforming in millimeter-wave multiuser {MISO} downlink,'' {\em IEEE Trans. Wirel. Commun.}, vol.~15, no.~2, pp.~1086--1100, Feb. 2016.

\bibitem{yu2017coverage}
X.~Yu, J.~Zhang, M.~Haenggi, and K.~B. Letaief, ``Coverage analysis for millimeter wave networks: The impact of directional antenna arrays,'' {\em IEEE J. Sel. Areas Commun.}, vol.~35, no.~7, pp.~1498--1512, 2017.

\bibitem{Jornet2011Channel}
J.~M. Jornet and I.~F. Akyildiz, ``Channel modeling and capacity analysis for electromagnetic wireless nanonetworks in the terahertz band,'' {\em IEEE Trans. Wirel. Commun.}, vol.~10, no.~10, pp.~3211--3221, Oct. 2011.

\bibitem{novlan2013analytical}
T.~D. Novlan, H.~S. Dhillon, and J.~G. Andrews, ``Analytical modeling of uplink cellular networks,'' {\em IEEE Trans. Wirel. Commun.}, vol.~12, no.~6, pp.~2669--2679, Jun. 2013.

\bibitem{Petrov2015Interference}
V.~Petrov, D.~Moltchanov, and Y.~Koucheryavy, ``Interference and sinr in dense terahertz networks,'' in {\em 2015 IEEE 82nd Vehicular Technology Conference (VTC2015-Fall)}, pp.~1--5, 2015.

\bibitem{singh2013offloading}
S.~Singh, H.~S. Dhillon, and J.~G. Andrews, ``Offloading in heterogeneous networks: Modeling, analysis, and design insights,'' {\em IEEE Trans. Wirel. Commun.}, vol.~12, no.~5, pp.~2484--2497, May. 2013.

\bibitem{chen2022deployment}
C.~Chen, J.~Zhang, X.~Chu, and J.~Zhang, ``On the deployment of small cells in {3D} {HetNets} with multi-antenna base stations,'' {\em IEEE Trans. Wirel. Commun.}, vol.~21, no.~11, pp.~9761--9774, Nov. 2022.

\bibitem{Zhang2020Energy}
H.~Zhang, H.~Zhang, W.~Liu, K.~Long, J.~Dong, and V.~C.~M. Leung, ``Energy efficient user clustering, hybrid precoding and power optimization in terahertz {MIMO-NOMA} systems,'' {\em IEEE J. Sel. Areas Commun.}, vol.~38, no.~9, pp.~2074--2085, Sept. 2020.

\bibitem{Andrews2011Tractable}
J.~G. Andrews, F.~Baccelli, and R.~K. Ganti, ``A tractable approach to coverage and rate in cellular networks,'' {\em IEEE Trans. Commun.}, vol.~59, no.~11, pp.~3122--3134, 2011.

\bibitem{alzer1997some}
H.~Alzer, ``On some inequalities for the incomplete gamma function,'' {\em Mathematics of Computation}, vol.~66, no.~218, pp.~771--778, 1997.

\bibitem{thornburg2016performance}
A.~Thornburg, T.~Bai, and R.~W. Heath, ``Performance analysis of outdoor {mmWave} {Ad Hoc} networks,'' {\em IEEE Trans. Signal Process.}, vol.~64, no.~15, pp.~4065--4079, Aug. 2016.

\end{thebibliography}
\end{document}